\documentclass[12pt,twoside]{article}
\usepackage[backend=bibtex,style=numeric-comp,autocite=plain,sorting=none,%
giveninits=true]{biblatex}
\usepackage{amsmath,amsfonts,amssymb,amsthm,mathabx}
\usepackage{times}
\usepackage{hyperref}
\hypersetup{
  breaklinks=true,   %
  colorlinks=true,   %
}
\usepackage{xurl}
\usepackage{tensor}
\usepackage{color}
\usepackage{caption}
\usepackage{enumitem}
\usepackage{graphicx}
\usepackage{array}
\usepackage{amsbsy}
\usepackage{appendix}
\usepackage{mathrsfs}
\usepackage{physics}
\usepackage{cancel}
\usepackage{yfonts}
\usepackage{inputenc}
\usepackage{chngcntr}
\usepackage{xcolor}
\usepackage{soul}
\usepackage{cancel}
\usepackage[leqno,fleqn,intlimits,newmultline]{empheq}

\advance\textheight\topskip
\numberwithin{equation}{section} \makeatletter
\@addtoreset{equation}{section}

\newcommand{\D}{d}

\newcommand{\p}{\partial}
\newcommand{\ts}{t}

\begin{document}

\title{Modular properties of massive scalar partition functions}

		\author{Ankit Aggarwal, Glenn Barnich}

    \def\mytitle{Modular properties of massive scalar partition functions}
    \pagestyle{myheadings} \markboth{\textsc{\small A.~Aggarwal, G.~Barnich}}
    {\textsc{\small Massive complex scalar field partition functions on flat
        backgrounds}}

\addtolength{\headsep}{4pt}

\begin{centering}

\vspace{1cm}

\textbf{\Large{\mytitle}}

\vspace{1.5cm}

{\large Ankit Aggarwal$ ^{a} $, Glenn Barnich$^b$}

\vspace{1cm}

\begin{minipage}{.9\textwidth}\small \it \begin{center}
    $^a$Institute for Theoretical Physics, TU Wien\\
    Wiedner Hauptstrasse 8–10/136, A-1040 Vienna, Austria\\
    E-mail:
\href{mailto:aggarwal@hep.itp.tuwien.ac.at}{aggarwal@hep.itp.tuwien.ac.at}
\end{center}
\end{minipage}

\vspace{.5cm}
\begin{minipage}{.9\textwidth}\small \it \begin{center}
	$^b$Physique Th\'eorique et Math\'ematique \\ Universit\'e libre de Bruxelles
	and International Solvay Institutes\\ Campus Plaine C.P. 231, B-1050
	Bruxelles, Belgium\\
  E-mail: \href{mailto:Glenn.Barnich@ulb.be}{Glenn.Barnich@ulb.be}
\end{center}
\end{minipage}

\vspace{.5cm}

\end{centering}

\vspace{1cm}

\begin{center}
  \begin{minipage}{.9\textwidth} \textsc{Abstract}. We compute the exact thermal
    partition functions of a massive scalar field on flat spacetime backgrounds
    of the form $\mathbb R^{d-q}\times \mathbb T^{q+1}$ and show that they possess an
    ${\rm SL}(q+1,\mathbb Z)$ symmetry. Non-trivial relations between equivalent
    expressions for the result are obtained by doing the computation using
    functional, canonical and worldline methods. For $q=1$, the results exhibit
    modular symmetry and may be expressed in terms of massive Maass-Jacobi
    forms. In the complex case with chemical potential for ${\rm U}(1)$ charge
    turned on, the usual discussion of relativistic Bose-Einstein condensation
    is modified by the presence of the small dimensions.
\end{minipage}
\end{center}

\vfill
\thispagestyle{empty}

\newpage

{\small
  \tableofcontents}

\section{Introduction}
\label{sec:introduction}

${\rm SL}(q+1,\mathbb Z)$ invariant Eisenstein series have been extensively
studied in the context of string theory scattering amplitudes (see e.g.
\cite{Obers:1999um,fleig_gustafsson_kleinschmidt_persson_2018}). Another,
arguably simpler, context where they naturally occur is as (the logarithm of)
thermal partition functions of a real massless scalar field on flat spacetime
backgrounds of the form $\mathbb{R}^{d-q}\times \mathbb{T}^{q+1}$ \cite{Alessio:2021krn}. For one small
spatial dimension, $q=1$, these partition functions are thus modular invariant.
This leads one directly to the question addressed in this follow-up paper: what
is the effect of mass on ${\rm SL}(q+1,\mathbb Z)$ invariance, and for $q=1$, on
modular invariance?

While there is a considerable amount of literature on how to compute such
partition functions and implications for relativistic Bose-Einstein condensation
(see e.g.
\cite{Hawking1977,Ambjorn:1981xw,Plunien:1987fr,doi:10.1142/2065,Kapusta:1981aa,%
  Haber:1981fg,Haber:1981ts}) on the one hand, and on (generalized) modular
invariance for conformal field theories in higher dimensions
\cite{Cappelli:1988vw,Cardy:1991kr,Dolan1998,Shaghoulian:2015kta,Shaghoulian:2016gol,%
  Horowitz:2017ifu,Luo2022} on the other, the precise breaking of this
invariance by a massive deformation and how it can be subsequently restored has
not been studied in detail (see however the recent related works
\cite{Downing2023} for massive free fermions on a torus, and
\cite{Karydas:2023ufs} for a massive complex scalar in odd spacetime dimensions,
$q=0,d=2L$). In this work, we revisit relativistic Bose-Einstein condensation on
$\mathbb{R}^{d-q}\times \mathbb{T}^{q+1}$ and generalized modular symmetry on the same background in
the presence of mass. We find the natural generalizations of
${\rm SL}(q+1,\mathbb Z)$ Eisenstein series in the massive case and show that
they agree with recently discovered massive deformations of Maass-Jacobi forms
\cite{Berg:2019jhh} when $q=1$.

We start in section \ref{sec:compl-mass-scal} by briefly reviewing how to deal
with toric backgrounds when computing partition functions. We closely follow
\cite{Alessio:2021krn} to which we refer for more details\footnote{In the
  published version of this reference, before equation (3.6), the change of
  variables should read $y=|n_{d+1}+n_d\tau|\sinh x$. Furthermore, the reflection
  formula (3.12) should read $\xi(z;\tau,\bar\tau) =\xi(1-z;\tau,\bar\tau)$.}. The skewed torus
$\mathbb{T}^{q+1}$ is the quotient of flat space by the lattice generated by constant
frame vectors. In terms of coordinates that make the torus square with unit
periodicities, the Euclidean second order action becomes that of a scalar field
coupled to a flat background metric determined by the frame vectors.

In order to relate to results obtained from the operator approach, we express
the Euclidean action in first order ADM form. This provides the identification
of the Hamiltonian together with the additional observables and their (purely
imaginary) chemical potentials. As is well known (see e.g.~\cite{Kapusta2006}),
this analysis also allows one to determine the modification of the second order
Euclidean action due to the inclusion of a chemical potential for ${\rm U}(1)$
charge in the case of a complex scalar field.

In section \ref{sec:comp-part-funct}, we compute the partition function, first
through the worldline formalism, then using functional/zeta function methods and
finally through operator quantization. The worldline formalism yields a
manifestly ${\rm SL}(q+1,\mathbb Z)$ invariant integral representation which
readily follows from the known heat kernels for a free particle on a real line
and a free particle on a unit circle. An alternative equivalent integral
expression is obtained by using the Poisson resummation formula. Whereas the
former expression is connected to the partition function computed in the
functional approach, the latter relates to the one coming from canonical
quantization, and is adapted to a low temperature/small box expansion. We end
the section by providing an expression adapted to a high temperature/large
volume expansion.

In section \ref{sec:particular-cases}, we particularize the results to the case
of a real massive scalar. In particular, we show that, for one small spatial
dimension, $q=1$, the logarithm of the partition function is essentially given
by massive Maass-Jacobi forms. Finally, in section \ref{sec:BEC}, we go back to
the complex charged case, and briefly discuss how Bose-Einstein condensation
works in the model along the lines of
\cite{Kapusta:1981aa,Haber:1981fg,Haber:1981ts,Singh1984}.

In a series of appendices, we collect various definitions and formulae, and
provide details on intermediate computations. We also make explicit a relation
(that we needed in \cite{Aggarwal:2022rrp}) between sums of Bessel functions and
express some of our results in terms of polylogarithms in order to connect to
the original derivations in the literature.

\section{Complex massive scalar field on
  $\mathbb{R}^{d-q}\times \mathbb{T}^{q+1}$}
\label{sec:compl-mass-scal}

In this section, we describe the setup we use for studying thermal partition
functions of a massive complex scalar field in a box with $ q $ small and $d-q$
large spatial dimensions with certain chemical potentials turned on. We discuss
the non-trivial part, $\mathbb{T}^{q+1}$, of the background in terms of adapted
coordinates and a constant frame and provide the action and Hamiltonian of the
scalar field as well as different conserved charges for which chemical
potentials are turned on.

\subsection{The skewed torus}

The large spatial dimensions are represented by $\mathbb R^p$, with $p=d-q$ and
the coordinates are denoted by $x^I$, $I=1,...,p$. The small dimensions,
including Euclidean time, are
described by the skewed torus $\mathbb T^{q+1}$. It is defined as the quotient of
$\mathbb{R}^{q+1}$ by a lattice $\Lambda^{d+1}$ that is generated by a set of
$q+1$ constant frame vectors ${ e}\indices{^a_\alpha}\in\mathbb{R}^{d+1}$:
\begin{equation}
\label{F423}
\mathbb{T}^{q+1}=\mathbb{R}^{q+1}/\Lambda^{q+1},\quad
\Lambda^{q+1}=\big\{m^{\alpha}{e}\indices{^a_\alpha}|\,
m^{\alpha}\in\mathbb{Z}^{q+1}\big\},
\end{equation}
with $\alpha=1,\dots,q+1$. The coordinates on $ \mathbb{R}^{q+1} $ are denoted by
$x^a=(x^i,x^{d+1})$, $i=p+1,\dots d$, with $ x^{d+1} $ parametrizing the
Euclidean-time circle. We consider a complex scalar field
$\Phi=\frac{1}{\sqrt 2}(\phi^1+i\phi^2)$. Results for a real massive scalar are obtained
from the formulas below by setting $\phi^2=0$, or equivalently,
$\Phi=\bar\Phi=\frac{1}{\sqrt 2}\phi^1$. The fields on $\mathbb{R}^{d-q}\times \mathbb{T}^{q+1}$ satisfy the
periodicity conditions
\begin{equation}
\label{eq:219}
\phi^A(x^I,x^a)=\phi^A(x^I,x^{a}+{e}\indices{^a_\alpha}),\quad A=1,2,\quad \forall\alpha.
\end{equation}

\subsection{Lagrangian action, Hamiltonian, and Noether charges}

In $x^{a}$ coordinates, the information about the geometry of the torus is
encoded in the skewed periodicities determined by the frame vectors
${e}\indices{^a_\alpha}$. The Lagrangian Euclidean action is
\begin{align}
\label{eq:219.a}
S^E_L[\Phi,\bar\Phi;m]= \int_{V_{d+1}} \D^{d+1}x\,
[\partial_I\bar \Phi\partial^I\Phi+ \delta^{ab}\partial_a\bar \Phi\partial_b\Phi+m^2\bar\Phi\Phi].
\end{align}
There is a global $U(1)\sim SO(2)$ symmetry under which the field transforms as
$\delta_\alpha\Phi=i\alpha \Phi$. The associated conserved Noether current is
\begin{equation}
  \label{eq:43}
  J^I=i(\bar \Phi\partial^I\Phi-\partial^I\bar \Phi\Phi),\quad J^a=i(\bar \Phi\partial^a\Phi-\partial^a\bar \Phi\Phi),
\end{equation}
with charge
\begin{equation}
  \label{eq:44}
  Q=i\int_{V_d} \D^{d}x (\bar \Phi \partial^{d+1}\Phi
  -\partial^{d+1}\bar \Phi\Phi)\\=-\int_{V_d} \D^dx\ \epsilon_{AB}\partial^{d+1}\phi^A\phi^B.
\end{equation}
with $\epsilon_{AB}$ completely antisymmetric and $\epsilon_{12}=1$. The other
Noether charges relevant for us here are the linear momenta in the small
directions,
\begin{equation}
  \label{eq:14}
  P_i=-\int_{V_d} \D^dx\, \partial^{d+1}\phi_A\partial_i\phi^A.
\end{equation}
In the canonical formalism, the Hamiltonian and Noether charges are given
by
\begin{equation}
  \label{eq:17}
  \begin{split}
    H&=\frac 12 \int_{V_d} \D^dx\, [\pi^A\pi_A+\partial_I\phi^A\partial^I\phi_A+\delta^{ij}\partial_i\phi^A\partial_j\phi_A+m^2\phi^A\phi_A],\\ P_i&=-\int_{V_d} \D^dx\, \pi^A\partial_i\phi_A,\quad
  Q=-\int_{V_d} \D^{d}x\, \epsilon_{AB}\pi^A\phi^B.
  \end{split}
\end{equation}

\subsection{Formulation in adapted coordinates with unit periodicities}

Associated to the $q+1$ linearly independent frame vectors, there is a co-frame
${e}\indices{_{a}^{\alpha}}$ such that
\begin{equation}
\label{eq:212}
{ e}\indices{_\alpha^a}{ e}\indices{_b^\alpha}=\delta^a_b,\quad
{ e}\indices{_\alpha^a}{ e}\indices{_a^\beta}=\delta^\beta_\alpha.
\end{equation}
This allows one to define coordinates in which periodicities of the field become
unity in all these directions separately,
\begin{equation}
\label{eq:220}
y^{\alpha}={ e}\indices{_a^{\alpha}} x^{a},\quad \phi(x^I,y^{\alpha})=\phi(x^I,y^{\alpha}+\delta_\beta^\alpha),\quad\forall \beta.
\end{equation}
In turn, the Cartesian metric $\delta_{ab}$ becomes
\begin{equation}
\label{eq:218}
g_{\alpha\beta}={e}\indices{^a_\alpha}\delta_{ab}{e}\indices{^b_\beta},
\end{equation}
When working in $y^\alpha$ coordinates, the information on the non-trivial
geometry of the torus gets encoded in the Lagrangian action through the coupling
of the scalar field to the flat, constant metric $ g_{\alpha\beta}$,
\begin{equation}
  \label{eq:Lcurved}
S^E_L[\phi; g,m]=\frac 12 \int_{V_{d+1}}
\D^px\D^{q+1}y\, \sqrt{ g}[g^{\alpha\beta}
\partial_\alpha\phi^A\partial_\beta \phi_A+\partial_I\phi^A\partial^I\phi_A+m^2\phi^A\phi_A].
\end{equation}
The volume of the torus is
$\text{V}_{q+1}=\mathrm{det}\, { e}\indices{^a_\alpha}=\sqrt{{g}}$. For later use, we
define the re-scaled frame, co-frame, metric and inverse metric,
\begin{equation}
\label{eq:224}
\begin{split}
\hat e\indices{^a_\alpha}=( \text{V}_{q+1})^{-\frac{1}{q+1}}
{e}\indices{^a_\alpha},\quad \hat e\indices{_a^\alpha}=
( \text{V}_{q+1})^{\frac{1}{q+1}}{e}\indices{_a^\alpha},\\
\hat g_{\alpha\beta}=
( \text{V}_{q+1})^{-\frac{2}{q+1}} g_{\alpha\beta},
\quad
\hat g^{\alpha\beta}=( \text{V}_{q+1})^{\frac{2}{q+1}} g^{\alpha\beta},
\end{split}
\end{equation}
all of which have unit determinants.

\subsection{First order formulation}
\label{sec:first-order-form}

In order to understand the operator content of the partition function, one needs
to work out the first order action principle associated to \eqref{eq:Lcurved},
which relies on the ADM decomposition \cite{Arnowitt:1962aa} adapted to
Euclidean signature of the flat metric,
\begin{equation}
\label{eq:214}
g^{\rm ADM}_{\alpha\beta}=
\begin{pmatrix}
h_{\iota\kappa} & N_\kappa\\ N_{\iota} & N^2+N_\lambda N^\lambda
\end{pmatrix},\quad
g^{\alpha\beta}_{\rm ADM}=
\begin{pmatrix}
h^{\iota\kappa} +\frac{N^\iota N^\kappa}{N^2}&
-\frac{N^\kappa}{N^2}\\ -\frac{N^{\iota}}{N^2}
& \frac{1}{N^2}
\end{pmatrix},
\end{equation}
where $\iota=d-q+1,\dots,d$ and $N_\iota= h_{\iota\kappa} N^\kappa$,
$\sqrt{g_{\rm ADM}}=N\sqrt h$, $\sqrt h=\text{V}_q$ and the associated moving frame may
be chosen as
\begin{equation}
  \label{eq:216}
  { e}\indices{_{\rm ADM}^{a}_\alpha}=
  \begin{pmatrix}
    \theta\indices{^i_\iota} & \theta\indices{^i_\kappa} N^\kappa \\
    0 \hdots 0 & N
  \end{pmatrix},\qquad
  { e}\indices{^{\rm ADM}_{a}^\alpha}=
  \begin{pmatrix}
    \theta\indices{_i^\iota} & \begin{matrix} 0\\ \vdots \\ 0
                             \end{matrix}\\
    -\frac{N^\iota}{N} & \frac{1}{N}
  \end{pmatrix},
\end{equation}
with $\theta\indices{^i_\iota}$ a moving frame for the spatial metric,
\begin{equation}
  \label{eq:215}
  h_{\iota\kappa}=\theta\indices{^i_\iota}
  \delta_{ij}\theta\indices{^j_\kappa},\quad {\rm det}\, {\theta^i}_\iota=\sqrt h=\text{V}_q.
\end{equation}

In these terms, the action to be used in the Hamiltonian path integral is
\begin{equation}
  \label{eq:52}
  S^E_H[\phi,\pi;g^{\rm ADM},m,\mu] = \int_{V_{d+1}} \D^px\D^{q+1}y\,
  \Big[-i\pi_A \partial_{d+1}\phi^A+N\mathcal H(h)+
  iN^\iota\mathcal H_\iota(h)],
\end{equation}
where
\begin{equation}
  \label{eq:46}
  \begin{split}
  \mathcal H(h,m) &=\frac{1}{2\sqrt{h}}
  \pi^A\pi_A +\frac 12 \sqrt{h}
  (h^{\iota\kappa}\partial_\iota\phi^A\partial_\kappa\phi_A
  +\partial_I\phi^A\partial^I\phi_A+m^2\phi^A\phi_A),\\
    \mathcal H_\iota(h)&=  \pi_A \partial_\iota \phi^A.
  \end{split}
\end{equation}
Defining
\begin{equation}
  \label{eq:2}
  \begin{split}
H(h,m)&=\int_{V_d} \D^px\D^d y\, \mathcal{H}(h,m),\\ H_{\iota}(h)&=\int_{V_d} \D^px\D^dy\,
\mathcal{H}_{\iota}(h)=-\theta\indices{^i_\iota} P_i,\quad
    P_i=-\int_{V_d} \D^dx\,\pi_A\partial_i\phi^A,
  \end{split}
\end{equation}
the Lagrangian partition function is equivalent to a phase space
path integral weighted by the first order Euclidean action \eqref{eq:52}. In
operator formalism, the latter corresponds to
\begin{equation}
\label{eq:4}
{Z}_{d,q}(\beta,\mu^j,h,m)
={\rm Tr}\, e^{-\beta [\hat H(h,m)-i\mu^j \hat  P_j]},
\end{equation}
with the identifications
\begin{equation}
\label{eq:5}
\beta=N,\quad \mu^j=\theta\indices{^j_\iota}\frac{N^\iota}{N}.
\end{equation}
It then follows from the expression of the determinant of the metric in ADM
parametrization that
\begin{equation}
  \label{eq:48}
  \text{V}_{q+1}=\beta \text{V}_q.
\end{equation}
We will also need,
\begin{equation}
  \label{eq:224a}
  \begin{split}
    \hat \theta\indices{^i_\iota}=(\text{V}_{q})^{-\frac{1}{q}}
    {\theta}\indices{^i_\iota},\quad \hat e\indices{_i^\iota}=
    (\text{V}_{q})^{\frac{1}{q}}{\theta}\indices{_i^\iota},\\
    \hat h_{\iota\kappa}=
    (\text{V}_{q})^{-\frac{2}{q}} h_{\iota\kappa},
    \quad
    \hat g^{\iota\kappa}=(\text{V}_{q})^{\frac{2}{q}} g^{\iota\kappa},
  \end{split}
\end{equation}
all of which have again unit determinants, as well as the dimensionless
temperature
\begin{equation}
  \label{eq:66}
  b=( \text{V}_q)^{-\frac{1}{q}}\beta.
\end{equation}

Whereas the above parametrization is useful in order to derive exact expressions
for the partition function adapted to a low temperature/small volume expansion
$b\gg1$, a dual parametrization of the ADM metric \cite{Alessio:2021krn} is
needed if one is interested in a high temperature/large volume expansion
$b\ll 1$:
\begin{equation}
\label{eq:239}
g_{\rm ADM}^{\alpha\beta}=
\begin{pmatrix}
h\indices{_D^{\iota\kappa}} & N^\kappa_D\\ N^{\iota}_D & N^2_D+N^D_\lambda N^\lambda_D
\end{pmatrix},\quad
g^{\rm ADM}_{\alpha\beta}=
\begin{pmatrix}
h\indices{^D_{\iota\kappa}} +\frac{N_\iota^D N^D_\kappa}{N^2_D}&
-\frac{N^D_\kappa}{N^2_D}\\ -\frac{N^D_{\iota}}{N^2_D}
& \frac{1}{N^2_D}
\end{pmatrix},
\end{equation}
with $N^\iota_D= h\indices{_D^{\iota\kappa}} N_\kappa^D$. This parametrization is
related to the standard one through
\begin{equation}
\label{eq:244}
\begin{split}
h\indices{_D^{\iota\kappa}}&=h\indices{^{\iota\kappa}}
+\frac{N^\iota N^\kappa}{N^2},\quad
\frac{1}{N_D^2}=N^2+N_\lambda N^\lambda,\quad  N^\iota_D=-\frac{N^\iota}{N^2},\\
h\indices{^D_{\iota\kappa}}&=h\indices{_{\iota\kappa}}
-\frac{N_\iota N_\kappa}{N^2+N_\lambda N^\lambda},
\end{split}
\end{equation}
with inverse relations
\begin{equation}
\label{eq:245}
\frac{1}{N^2}=N^2_D+N^D_\lambda N^\lambda_D,
\quad N_\iota=-\frac{N_{\iota}^D}{N_D^2},\quad
h\indices{_{\iota\kappa}}=h\indices{^D_{\iota\kappa}}
+\frac{N_\iota^D N^D_\kappa}{N^2_D}.
\end{equation}
Computing determinants, one finds
\begin{align}
\label{eq:245.1}
  \text{V}_q^D= \text{V}_q\frac{N}{\sqrt{N^2+N_{\lambda}N^{\lambda}}}=\text{V}_q\frac{1}{\sqrt{1+\mu_i\mu^i}},\quad b_D
  = N_D ( \text{V}^D_q)^{\frac{1}{q}}=\frac{1}{b}\frac{1}{(1+\mu_i\mu^i)^{\frac{q+1}{2q}}}~.
\end{align}

\subsection{Turning on ${\rm U}(1)$ charge}
\label{sec:rm-u1-charge}

Turning on an additional $U(1)$ charge, denoted by $Q$ and defined in equation
\eqref{eq:17}, amounts to computing the partition function
\begin{empheq}[box=\fbox]{equation}
  \label{eq:3}
  {Z}_{d,q}(\beta,\mu^j,\theta\indices{_i^\iota},m,\nu)
  ={\rm Tr}\, e^{-\beta [\hat H(h,m)-i\mu^j \hat  P_j-\nu\hat Q]}~.
\end{empheq}
Here $ \nu $ is the chemical potential associated with $ \hat Q $. This partition
function gives rise to a phase space path integral where the first order action
\eqref{eq:52} contains the additional term
\begin{equation}
  \label{eq:18}
  \int_{V_{d+1}} \D^px\D^{q+1}y\,N \epsilon_{AB} \nu \pi^A\phi^B.
\end{equation}
When integrating over the momenta, one finds that the
extremum  for this extended action is given by
\begin{equation}
  \label{eq:50}
  \pi^A=\sqrt h(i { e}\indices{^{\rm ADM}_{d+1}^\alpha} \partial_\alpha\phi^A-\nu\epsilon^{AB}\phi_B).
\end{equation}
The Lagrangian action obtained after the integration is
\begin{multline}
  \label{eq:56a}
  S^E_L[\phi;e,m,\nu]=\frac 12 \int_{V_{d+1}} \D^px
  \D^{q+1}y\sqrt g \Big[ \partial_I\phi^A\partial^I\phi_A+(m^2-\nu^2)\phi^A\phi_A\\
  + g^{\alpha\beta}\partial_\alpha\phi^A\partial_\beta\phi_A +2i\nu {e}\indices{_{d+1}^\alpha}\partial_\alpha\phi^A\epsilon_{AB}\phi^B\Big],
\end{multline}
with the understanding that the vielbein ${ e}\indices{_{a}^\alpha}$ and metric
$g^{\alpha\beta}$ are written in ADM parametrization.

\section{Representations of the partition function}
\label{sec:comp-part-funct}

Our aim in the following section is to compute and study the properties of the partition
function
\begin{empheq}[box=\fbox]{equation}
  \label{eq:217}
  Z_{d,q}(e,m,\nu)=\int \mathcal D\phi
  \, e^{-S^E_L[\phi;e,m,\nu]}.
\end{empheq}
This can be done in any parametrization; the ADM parametrization is only needed
in order to interpret the result in the operator approach as in \eqref{eq:3}.

Since the Lagrangian is quadratic,
$S^E_L=-\frac 12 \phi^\Gamma\mathcal D_{\Gamma\Delta}\phi^\Delta$, the path
integral \eqref{eq:217} is formally given by a functional determinant
\begin{equation}
  \label{eq:53}
  \ln Z_{d,q}(e,m,\nu)=-\frac 12 \ln {\rm Det}\, \mathcal D_{\Gamma\Delta}
  =-\frac 12 {\rm Tr}\ln \mathcal D_{\Gamma\Delta}.
\end{equation}
\subsection{${\rm SL}(q+1,\mathbb Z)$ invariance of the partition function}

As in the massless case, the partition function of a massive real scalar field
may be expected to be invariant under SL$(q+1,\mathbb Z)$ transformations of the
lattice frame vectors
\begin{equation}
	\label{eq:15}
	e\indices{^\prime^a_\alpha}=S\indices{_\alpha^\beta} e\indices{^a_\beta},\quad S\indices{_\alpha^\beta}\in{\rm SL}(q+1,\mathbb Z).
\end{equation}
This invariance can be seen by the following argument which is same as for the massless case in \cite{Alessio:2021krn}, section 4.2. The action \eqref{eq:56a} is diffeomorphism invariant where derivatives, fields, metric, and vielbeins are transformed appropriately under diffeomorphisms. In particular, the mass term is also invariant by itself. However, this invariance is broken to down to
SL$(q+1,\mathbb Z)$ transformations due to the periodic boundary conditions in $y^\alpha$. Assuming that the path integral measure remains invariant, this implies that the partition function is invariant
\begin{equation}
	Z_{d,q}[e,m,\nu]=Z_{d,q}[e',m,\nu],
\end{equation}
with ${\rm det}\,g'_{\alpha\beta}={\rm det}\,g_{\alpha\beta}$. More details can be found in
\cite{Alessio:2021krn}.
\subsection{Worldline approach}
\label{sec:worldline-approach-1}

In the worldline formalism, the trace in equation \eqref{eq:53} can be
represented as an integral over Schwinger proper time, $ t $. In these terms,
the partition function admits the representation
\begin{equation}
\label{eq:49}
\begin{split}
& \ln Z_{d,q}(e,m,\nu)=\frac 12 \int^\infty_0\frac{d\ts}{\ts}\ {\rm Tr}\
e^{-\ts\hat H^P},\\ & \hat H^P=(\hat p_I\hat p^I+\hat p_\alpha g^{\alpha\beta}
\hat p_\beta+m^2-\nu^2)\delta_{AB} +2\nu e\indices{_{d+1}^\alpha}\hat p_\alpha \epsilon_{AB},
\end{split}
\end{equation}
where one has to properly remove zero modes. The first quantized Hamiltonian
$\hat H^P$ acts in the tensor product of a free particle Hilbert space with a
2-dimensional internal space in order to take into account the real and the
imaginary part of the scalar field as well as the ${\rm U}(1)$ charge.

The associated heat kernel,
\begin{equation}
  \label{eq:20}
  K_{\mathbb{R}^p\times\mathbb{T}^{q+1}}(x'^I,y'^\alpha,x^I,y^\alpha;\ts)_{AB}= \langle x'^I,y'^\alpha,A|e^{-\ts\hat H^P}|x^I,y^\alpha,B\rangle
\end{equation}
and partition function are worked out in Appendix \ref{sec:comp-wordl-form} in
terms of the redefined parameters,
\begin{equation}
    \label{eq:59}
    \rho_L=\frac{m}{2\pi}( \text{V}_{q+1})^{\frac{1}{q+1}},\quad \nu_L=\frac{\nu}{2\pi}( \text{V}_{q+1})^{\frac{1}{q+1}},\quad
    w^\pm_\alpha=\mp i\nu_{L} { \hat e}\indices{^{d+1}_\alpha}.
\end{equation}
After a suitable subtraction, one finds that the partition function admits the
following integral representation in terms of the Riemannn theta function
defined in \eqref{F5},
\begin{empheq}[box=\fbox]{equation}
  \label{eq:59b}
  \ln Z_{d,q}(e,m,\nu)
    =\frac{V_{d-q}}{2( \text{V}_{q+1})^{\frac{d-q}{q+1}}}
      \int_0^{\infty}\D\ts\,\ts^{\frac{d+1}{2}-1}e^{-\pi \frac{{\rho_L^2}}\ts}\sum_\pm\big(
      \vartheta_{q+1}(w^{\pm}| it \hat g)-1\big).
\end{empheq}

This expression is manifestly invariant under the ${\rm SL}(q+1,\mathbb Z)$
transformations \eqref{eq:15} that only affects the lattice vectors
$e\indices{^a_\alpha}$,
\begin{equation}
  \label{eq:29}
  \ln Z_{d,q}(e',m,\nu)=\ln Z_{d,q}(e,m,\nu).
\end{equation}
Indeed, the Riemann theta function is invariant since,
\begin{equation}
  m^{\alpha}\hat g'_{\alpha\beta}m^{\beta}=	m^{\alpha}S\indices{_\alpha^\gamma} \hat g_{\gamma\delta}
  S\indices{_{\beta}^{\delta}}m^{\beta}=m'^{\alpha}\hat g_{\alpha\beta}m'^{\beta},\quad w'^\pm_\alpha m^\alpha=w^\pm_\beta
  S\indices{_\alpha^\beta} m^\alpha=w^\pm_\alpha m'^\alpha,
\end{equation}
with $m'^\alpha=S\indices{_{\beta}^{\alpha}}m^{\beta}$. Hence, the transformation
amounts to a relabeling of the lattice points and, since the Riemann theta
function involves a sum over the entire $\mathbb{Z}^{q+1}$ lattice, it is
invariant under SL$(q+1,\mathbb{Z})$. Furthermore, both $\text{V}_{q+1} $ and
$\rho_L$ are invariant because the volume is preserved.

The subtraction, discussed in detail in Appendix \ref{sec:comp-wordl-form},
corresponds to the $-1$ in the formula above and removes the $m^\alpha=0$ zero
mode of the Riemann theta function. The associated integral \eqref{eq:67}
diverges in odd spatial dimensions. However we choose to subtract it in even
dimensions as well, without affecting ${\rm SL}(q+1,\mathbb Z)$ invariance.

\subsection{Functional approach}
\label{sec:functional-approach-2}

In the functional approach to computing partition functions
\cite{Dowker:1975tf,DeWitt:1975ys,Hawking:1976ja,Kapusta:1981aa} (see
e.g.~\cite{doi:10.1142/2065} for a review), one associates a zeta function to
the eigenvalues of the quadratic kernel of the Lagrangian action. In our case,
an orthonormal basis of eigenfunctions is given by
\begin{equation}
\label{FF1}
f_{n_I,n_{\alpha}}(x,y)=\frac{1}{\sqrt{V_{d+1}}}e^{ik_I x^I+2\pi
in_{\alpha} y^\alpha},
\end{equation}
with $k_I=\frac{2\pi n_I}{L_I}$, $n_I\in\mathbb{Z}^p$,
$n_{\alpha}\in\mathbb{Z}^{q+1}$, and $V_{d+1}=V_{d-q} \text{V}_{q+1}$.

The quadratic kernel in the path integral is determined by the operator
\begin{equation}
\label{eq:57}
\mathcal D_{AB}=[-\Delta +m^2-\nu^2]\delta_{AB}
-2i\nu e\indices{_{d+1}^\alpha}\partial_\alpha \epsilon_{AB},
\end{equation}
where $\Delta
=\partial_I\partial^I+ g^{\alpha\beta}\partial_\alpha\partial_\beta$ is the Laplacian
and the eigenvalue matrix, $ (\lambda^2_{n_I,n_\alpha})_{AB} $, is defined through
\begin{equation}
  \label{eq:51}
  \mathcal D_{AB}\,
  f_{n_I,n_{\alpha}}=(\lambda^2_{n_I,n_{\alpha}})_{AB}\,f_{n_I,n_{\alpha}}.
\end{equation}
The associated zeta functions $\zeta_{\mathcal D}(s)$ and partition functions
\begin{equation}
  \label{eq:71}
  \ln Z_{d,q}(e,m,\nu)=\frac 12 \zeta_{\mathcal D}'(0),
\end{equation}
are worked out in Appendix \ref{sec:comp-funct-appr} for $p>0$. The manifestly
${\rm SL}(q+1,\mathbb Z)$ invariant result is
\begin{empheq}[box=\fbox]{multline}
  \label{eq:81}
  \ln Z_{d,q}(e,m,\nu)=\frac{V_{d-q}}{(\text{V}_{q+1})^{\frac{d-q}{q+1}}}
  \Big[ \\
  2\sideset{}{'}\sum_{m^\alpha\in \mathbb Z^{q+1}}\big(\frac{\rho_L}{\sqrt{m^\alpha\hat g_{\alpha\beta}m^\beta}}
  \big)^{\frac{d+1}{2}}
  K_{\frac{d+1}{2}}\big(
  2\pi \rho_L \sqrt{m^\alpha\hat
    g_{\alpha\beta}m^\beta}\big)\cosh{(2\pi \nu_L{\hat e}\indices{^{d+1}_\alpha}m^\alpha)}\Big],
\end{empheq}
where the prime on the sum indicates that the term with $m^\alpha=0$ is excluded.

\subsection{Canonical approach}
\label{sec:canonical-approach-2}

We now want to directly evaluate \eqref{eq:3} in the operator approach. An
orthonormal basis of eigenfunctions is now
\begin{equation}
  \label{FF1a}
  f_{n_{I},n_{\iota}}(x)=\frac{1}{\sqrt{V_{d}}}e^{ik_Ix^I}e^{2\pi in_{\iota}\theta\indices{_i^\iota}x^i},
  \qquad k_I=\frac{2\pi n_I}{L_I},\qquad n_I\in\mathbb{Z}^p,\qquad n_{\iota}\in\mathbb{Z}^{q},
\end{equation}
where $V_{d}=V_{d-q}\text{V}_{q}$ and the orthonormality condition is
\begin{multline}
  \label{FF2a}
  (f_{n_{I},n_{\iota}},f_{n'_{I},n'_{\iota}})=\int_{V_{d}}
  \text{d}^{d}x\,f^{*}_{n_{I},n_{\iota}}(x)f_{n'_{I},n'_{\iota}}
  (x)  \\=\frac{1}{V_d}\int_{V_{d-q}}
  \text{d}^px\text{d}^qy\sqrt{h}\,e^{i(k'_I-k_I)x^I}e^{2\pi
    i(n'_{\iota}-n_{\iota})y^{\iota}}
  =\prod_{I,\iota}\delta_{n_I,n'_I}\delta_{n_{\iota},n'_{\iota}}.
\end{multline}
The fields and their momenta may be expanded as
\begin{equation}
  \label{FF4}
  \phi^A(x)=\sum_{(n_I,n_{\iota})\in\mathbb{Z}^d}
  \phi^A_{n_I,n_{\iota}}f_{n_{I},n_{\iota}}(x),\quad
  \pi_A(x)=\sum_{(n_I,n_{\iota})\in\mathbb{Z}^d}
  \pi_{An_I,n_{\iota}}f_{n_{I},n_{\iota}}(x),
\end{equation}
Defining the oscillator variables,
\begin{equation}
  \label{eq:11}
  a^A_{n_I,n_{\iota}}=\sqrt{\frac{\omega_{n_I,n_{\iota}}}{2}}\big[\phi^A_{n_I,n_{\iota}}+\frac{i}{\omega_{n_I,n_{\iota}}}\pi^A_{n_I,n_{\iota}}\big],\quad
  \omega_{n_I,n_{\iota}}=\sqrt{k_Ik^I+(2\pi)^2n_{\iota}h^{\iota\kappa}n_{\kappa}+m^2},
\end{equation}
these expansions become
\begin{equation}
  \begin{split}
  \label{eq:140z}
    \phi^A(x)&=\sum_{(n_I,n_{\iota})\in\mathbb{Z}^d}
              \frac{1}{\sqrt{2\omega_{n_I,n_{\iota}}}}(a^A_{n_I,n_{\iota}}f_{n_{I},n_{\iota}}(x)+{\rm c.c.}),\\
    \pi_A(x)&=-i\sum_{(n_I,n_{\iota})\in\mathbb{Z}^d}
             \sqrt{\frac{\omega_{n_I,n_{\iota}}}{2}}(a_{A,{n_I,n_{\iota}}}f_{n_{I},n_{\iota}}(x)-{\rm c.c.}).
  \end{split}
\end{equation}
Contrary to the massless case treated in \cite{Alessio:2021krn}, the term with
$(n_I,n_{\iota})=(0,\dots ,0,0,\dots ,0)$ can be included because the associated
angular frequency $\omega_0=m$ does no longer vanish.

Inserting this mode expansion into $H(h,m)$ and $P_i$, one finds for the
quantized Hamiltonian, the linear momenta in the small dimensions and the $U(1)$
charge,
\begin{equation}
  \label{FF9}
  \begin{split}
    \hat{H}(h) &= \sum_{(n_I,n_{\iota})\in\mathbb{Z}^d}
    \omega_{n_I,n_{\iota}}\hat{a}^{A\dagger}_{n_I,n_{\iota}}\hat{a}_{An_I,n_{\iota}}
    +2E^{d,q}_0(h,m),\quad \hat Q=-i\epsilon_{AB}\sum_{n_I,n_\iota\in \mathbb Z^q}a^{A\dagger}_{n_I,n_\iota}a^B_{n_I,n_\iota}, \\
    \hat{P}_i &=\theta_i{}^{\iota}
    \sum_{(n_i,n_{\iota})\in\mathbb{Z}^d}(2\pi i n_{\iota})
    \hat{a}^{A\dagger}_{n_I,n_{\iota}}\hat{a}_{An_I,n_{\iota}},
  \end{split}
\end{equation}
where $E^{d,q}_0(h,m)$ is the renormalized Casimir energy of a single massive
scalar field on $\mathbb{R}^p\times\mathbb{T}^q$. Note that, contrary to the
massless case, the contribution to the Hamiltonian of the degree of freedom
associated to $(n_I,n_{\iota})=(0,\dots ,0,0,\dots ,0)$ is not that of a free
particle but that of another oscillator with angular frequency $\omega_0=m$.

The result for the partition function is worked out in Appendix
\ref{sec:compcan}. In terms of the dimensionless parameter,
\begin{equation}
  \label{eq:13}
  \rho_H=\frac{m}{2\pi}( \text{V}_q)^{\frac{1}{q}}=\rho_Lb^{-\frac{1}{q+1}},
\end{equation}
one finds
\begin{empheq}[box=\fbox]{multline}
  \label{eq:37a}
  \ln Z_{d,q}(\beta,\mu^j,\theta\indices{_i^\iota},m,\nu)=\frac{V_{d-q}}{(\text{V}_{q})^{\frac{d-q}{q}}}
    \Big[\\
    2b\sideset{}{'}\sum_{m^\iota\in \mathbb Z^{q}}
    \Big(\frac{\rho_H}{\sqrt{m^\iota\hat h_{\iota\kappa}m^\kappa}}
    \Big)^{\frac{d+1}{2}}
    K_{\frac{d+1}{2}}\Big(
      2\pi \rho_H\sqrt{m^\iota\hat h_{\iota\kappa}m^\kappa}\Big) \\
    +4b^{-\frac{d-q-1}{2}}
    \sum_{l\in \mathbb{N}^*}\sum_{n_\iota\in
      \mathbb Z^q}\Big(\frac{\sqrt{n_\iota\hat
          h^{\iota\kappa}n_\kappa+\rho_H^2}}{l}\Big)^{\frac{d+1-q}{2}}
    K_{\frac{d+1-q}{2}}\Big(2\pi l b \sqrt{n_\iota\hat
        h^{\iota\kappa}n_\kappa+\rho_H^2}\Big)\\e^{2\pi i l n_\iota N^\iota}\cosh{(l\beta\nu)}
    \Big],
\end{empheq}
where the terms  on the second line correspond to $-2\beta E^{d,q}_0(h,m)$,
with
\begin{empheq}[box=\fbox]{equation}
    \label{eq:74}
    E_0^{d,q}(h,m)=-
    \frac{V_{d-q}}{(\text{V}_{q})^{\frac{d+1-q}{q}}}\Big[
    \sideset{}{'}\sum_{m^\iota\in \mathbb Z^{q}}
    \Big(\frac{\rho_H}{\sqrt{m^\iota\hat h_{\iota\kappa}m^\kappa}}
    \Big)^{\frac{d+1}{2}}
    K_{\frac{d+1}{2}}\Big(
    2\pi \rho_H\sqrt{m^\iota\hat h_{\iota\kappa}m^\kappa}\Big)
    \Big].
\end{empheq}

When taking into account that the $b$-dependent Bessel functions on the third
line are exponentially suppressed, cf. equation \eqref{eq:78}, the expression
for the partition function in \eqref{eq:37a} is directly adapted to a
low-temperature/small volume expansion $b\gg 1$ when the chemical potential
$\nu$ vanishes. When $\nu$ doesn't vanish, the expression is still adapted to a
low temperature expansion, but the term with $n_\iota=\{0,\dots,0\} $ is no
longer exponentially suppressed. This will be analyzed in detail in the
discussion on Bose-Einstein condensation in section \ref{sec:BEC} below.

\subsection{High temperature expansion}
\label{sec:high-temp-volume complex}

For a high temperature/large volume expansion $b\ll 1$. Here, one uses the dual
parametrization of the ADM metric described in \ref{sec:first-order-form} and
starts from expression \eqref{eq:59a} in the worldline approach. Defining
\begin{equation}
  \label{eq:1}
  \rho_H^D=\rho_H(bb_D)^{\frac{1}{q+1}}= \rho_H(1+\mu_i\mu^i)^{-\frac{1}{2q}}.
\end{equation}
After a suitable use of Poisson resummation, and the same subtraction discussed
previously, one ends up with
\begin{empheq}[box=\fbox]{multline}
  \label{eq:37c}
  \ln Z_{d,q}(\beta,\mu^j,\theta\indices{_i^\iota},m,\nu)=\frac{\text{V}_{d-q}}{( \text{V}_{q}^D)^{\frac{d-q}{q}}}
  \Big[\\
  4b_D^d\sum_{l\in \mathbb{N}^*}\cosh{(l\beta\nu)}\Big (\frac{\rho_H^D}{l b_D}\Big)^{\frac{d+1}{2}}
    K_{\frac{d+1}{2}}\Big (2\pi l \frac{\rho_H^D}{b_D}\Big )\\
  +b_D^{\frac{d}{2}}\sum_\pm\sideset{}{'}\sum_{m^\iota\in
      \mathbb Z^q}\sum_{n\in \mathbb
      Z}
    \Big(\frac{\sqrt{(\frac{\rho_H^D}{b_D})^2+(n\pm i \frac{\beta\nu}{2\pi})^2}}{\sqrt{m^\iota\hat
          h^D_{\iota\kappa}m^\kappa}}\Big)^\frac{d}{2}
      \\K_{\frac{d}{2}}\Big (2\pi b_D
      \sqrt{(\frac{\rho_H^D}{b_D})^2+(n\pm i \frac{\beta\nu}{2\pi})^2}\sqrt{m^\iota\hat
        h^D_{\iota\kappa}m^\kappa}\Big)e^{2\pi im^\iota N_\iota^D n\mp m^\iota N_\iota^D\beta\nu}
  \Big].
\end{empheq}

\section{Real massive scalar field}
\label{sec:particular-cases}

In this section, we discuss important special cases related to the real massive
scalar field. When $q=1$, we recall how ${\rm SL}(2,\mathbb Z)$ invariance is
related to modular invariance and express the partition function in terms of
massive Maass-Jacobi forms. The $ q=0 $ case corresponds to the blackbody result
for a massive scalar field.

The results are obtained by setting $\nu=0$, dividing by $2$ in terms that do
not involve the sum $\sum_{\pm}$, and dropping this sum in the others in the
relevant equations that we will highlight below.

\subsection{Arbitrary small and large dimensions}
\label{sec:arbitr-small-large}

From the worldline approach, using equations \eqref{eq:59b} one gets
\begin{equation}
\label{eq:17 a}
\ln {Z}_{d,q}(g,m)=\frac{V_{d-q}}{2( \text{V}_{q+1})^{\frac{d-q}{q+1}}}
\int_0^{\infty}\D\ts\,\ts^{\frac{d+1}{2}-1}e^{-\frac{\pi \rho_L^2}{\ts}}\Big(\theta_{q+1}(\ts\hat g)-1\Big)
\end{equation}
where $ \theta_{q+1}(t\hat{g})=\vartheta_{q+1}(0|t\hat g) $ is explicitly defined
in \eqref{F6}.

The expression from the functional approach using \eqref{eq:81} is
\begin{equation}
\label{eq:20 a}
\ln {Z}_{d,q}(g,m)=\frac{V_{d-q}}{( \text{V}_{q+1})^{\frac{d-q}{q+1}}}
\sideset{}{'}\sum_{m^\alpha\in \mathbb Z^{q+1}}
\Big(\frac{\rho_L}{\sqrt{m^\alpha\hat g_{\alpha\beta}m^\beta}}
\Big)^{\frac{d+1}{2}}
K_{\frac{d+1}{2}}\Big(
2\pi \rho_L\sqrt{m^\alpha\hat g_{\alpha\beta}m^\beta}\Big).
\end{equation}
Using \eqref{BesselK limit}, one recovers the result of \cite{Alessio:2021krn}
in the massless limit $\rho_L\to 0^+$ in terms of a real analytic Eisenstein
series,
\begin{equation}
\label{eq:23}
\ln {Z}_{d,q}(g)=\frac{V_{d-q}}{( \text{V}_{q+1})^{\frac{d-q}{q+1}}}
\frac{\Gamma(\frac{d+1}{2})}{2\pi^{\frac{d+1}{2}}}\sideset{}{'}\sum_{m^\alpha\in \mathbb Z^{q+1}}
\left(m^\alpha\hat g_{\alpha\beta}m^\beta
\right)^{-\frac{d+1}{2}}.
\end{equation}

From the canonical approach, one obtains using \eqref{eq:37a},
\begin{multline}
  \label{eq:37d}
  \ln Z_{d,q}(\beta,\mu^j,h,m)=\frac{V_{d-q}}{(\text{V}_{q})^{\frac{d-q}{q}}}
  \Big[
  b\sideset{}{'}\sum_{m^\iota\in \mathbb Z^{q}}
  \Big(\frac{\rho_H}{\sqrt{m^\iota\hat h_{\iota\kappa}m^\kappa}}
  \Big)^{\frac{d+1}{2}}
  K_{\frac{d+1}{2}}\Big(
  2\pi \rho_H\sqrt{m^\iota\hat h_{\iota\kappa}m^\kappa}\Big)+ \\
  2b^{-\frac{d-q-1}{2}}
  \sum_{l\in \mathbb{N}^*}\sum_{n_\iota\in
    \mathbb Z^q}\Big(\frac{\sqrt{n_\iota\hat
      h^{\iota\kappa}n_\kappa+\rho_H^2}}{l}\Big)^{\frac{d+1-q}{2}}
  K_{\frac{d+1-q}{2}}\Big(2\pi l b \sqrt{n_\iota\hat
    h^{\iota\kappa}n_\kappa+\rho_H^2}\Big)e^{2\pi i l n_\iota N^\iota}
  \Big],
\end{multline}
and associated Casimir energy given in \eqref{eq:74}.
In the massless limit $\rho_H\to 0^+$, this reduces to
\begin{multline}
  \label{eq:41}
  \ln{Z}_{d,q}(\beta,\mu^j,h)=
  \frac{V_{d-q}}{( \text{V}_q)^{\frac{d-q}{q}}}
  \bigg[b\frac{\Gamma(\frac{d+1}{2})}{2\pi^{\frac{d+1}{2}}}
  \sideset{}{'}\sum_{m^\iota\in \mathbb Z^{q}}
  \left(m^\iota\hat h_{\iota\kappa}m^\kappa
  \right)^{-\frac{d+1}{2}}+\frac{\Gamma(\frac{d+1-q}{2})\zeta(d+1-q)}{\pi^{\frac{d+1-q}{2}}}b^{-(d-q)}\\
  +2b^{-\frac{d-q-1}{2}}\sideset{}{'}\sum_{n_{\iota}\in\mathbb{Z}^q}
  \sum_{l\in\mathbb{N}^*}\bigg(\frac{\sqrt{n_{\iota}\hat h^{\iota\kappa}
      n_{\kappa}}}{l}\bigg)^{\frac{d+1-q}{2}}
  K_{\frac{d+1-q}{2}}(2\pi lb\sqrt{n_{\iota} \hat h^{\iota\kappa}n_{\kappa}})
  e^{2\pi i l n_{\iota} N^{\iota}}\bigg],
\end{multline}
with Casimir energy
\begin{equation}
  \label{eq:7}
  E_0^{d,q}(h)=-
  \frac{V_{d-q}}{( \text{V}_{q})^{\frac{d+1-q}{q}}}
  \frac{\Gamma(\frac{d+1}{2})}{2\pi^{\frac{d+1}{2}}}
  \sideset{}{'}\sum_{m^\iota\in \mathbb Z^{q}}
  \left(m^\iota\hat g_{\iota\kappa}m^\kappa
  \right)^{-\frac{d+1}{2}}.
\end{equation}

The expression for the partition function adapted to a high temperature/large
volume expansion is obtained using \eqref{eq:37c}
\begin{multline}
  \label{eq:35}
  \ln Z_{d,q}(g,m)=\frac{V_{d-q}}{( \text{V}^D_{q})^{\frac{d-q}{q}}}
  \Big[2b_D^d\sum_{l\in \mathbb{N}^*}\Big(\frac{\rho_H^D}{lb_D}\Big)^{\frac{d+1}{2}}
  K_{\frac{d+1}{2}}\Big(2\pi l\frac{\rho_H^D}{b_D}\Big)\\
  +b_D^{\frac{d}{2}}   \sideset{}{'}\sum_{m^\iota\in \mathbb Z^{q}}\sum_{n\in\mathbb
    Z} \Big(\frac{\sqrt{n^2+(\frac{\rho_H^D}{b_D})^2}}{\sqrt{m^\iota\hat{h}_{\iota\kappa}^Dm^\kappa}}
  \Big)^{\frac{d}{2}}
  K_{\frac{d}{2}}\Big(2\pi b_D \sqrt{n^2+(\frac{\rho_H^D}{b_D})^2} \sqrt{m^\iota\hat
      h_{\iota\kappa}^Dm^\kappa}\Big)e^{2\pi i n m^{\iota}N_{\iota}^D}
  \Big].
\end{multline}

In the massless limit $\rho_H^D\to 0^+$, one finds
\begin{multline}
\label{eq:39}
\ln Z_{d,q}(g,m)=\frac{V_{d-q}}{(\text{V}^D_{q})^{\frac{d-q}{q}}}
\Big[
b_D^d\frac{\Gamma(\frac{d+1}{2})\zeta(d+1)}{\pi^{\frac{d+1}{2}}}+\frac{\Gamma(\frac{d}{2})}{2\pi^{\frac{d}{2}}}
\sideset{}{'}\sum_{m^\iota\in \mathbb Z^{q}}({m^\iota\hat{h}_{\iota\kappa}^Dm^\kappa})^{-\frac{d}{2}}\\
+2b_D^{\frac{d}{2}}
\sum_{n\in\mathbb
	N^*}  \sideset{}{'}\sum_{m^\iota\in \mathbb Z^{q}}
\Big(\frac{n}{\sqrt{m^\iota\hat{h}_{\iota\kappa}^Dm^\kappa}}
\Big)^{\frac{d}{2}}
K_{\frac{d}{2}}\Big(2\pi n b_D\sqrt{m^\iota\hat
	h_{\iota\kappa}^Dm^\kappa}\Big)e^{2\pi i n m^{\iota}N_{\iota}^D}
\Big].
\end{multline}

\subsection{One small spatial dimension: massive deformations of Maass-Jacobi
  forms}
\label{sec:one-small-spatial-1}

In the case where the spacetime manifold is $\mathbb R^{d-1}\times \mathbb T^2$,
$q=1$,
\begin{equation}
  \label{eq:22}
  \text{V}_{2}=L_d \beta,\quad\text{V}_{1}=L_d,\quad
  \rho_L=\frac{m \sqrt{L_d\beta}}{2\pi},\quad b=\frac{\beta}{L_d},\quad
  \rho_H=\frac{mL_d}{2\pi}.
\end{equation}
with $L_d$ the length
of the small spatial dimension.

The modular parameter is defined in terms of the frame vectors as \begin{equation}
  \label{eq:30}
  \tau=
  \frac{e\indices{^d_{d+1}}+ie\indices{^{d+1}_{d+1}}}{e\indices{^d_{d}}+ie\indices{^{d+1}_{d}}}.
\end{equation}
An ${\rm SL}(2,\mathbb Z)$ transformation of the frame vectors as in
\eqref{eq:15}, parametrized as
\begin{equation}
  \label{eq:16}
  S\indices{_\alpha^\beta}=\begin{pmatrix} d & c\\ b & a
                           \end{pmatrix},\quad ad-bc=1
\end{equation}
induces the modular transformations
\begin{equation}
                           \label{eq:19}
                           \tau'=\frac{a\tau+b}{c\tau +d}.
\end{equation}
With this parametrization,
\begin{equation}
  \label{eq:21}
  \hat g_{\alpha\beta}=\frac{1}{\tau_2}\begin{pmatrix} 1  & \tau_1\\
           \tau_1 & \abs{\tau}^2
       \end{pmatrix},\quad
       m^\alpha\hat g_{\alpha\beta}m^\beta=\frac{1}{\tau_2}|r\tau+l|^2,\quad m^1=l,\, m^2=r,
       \end{equation}
       and the ${\rm SL}(2,\mathbb Z)$ invariant partition function becomes
\begin{equation}
  \label{eq:17e}
  \begin{split}
  \ln {Z}_{d,1}(g,m)&=\frac{V_{d-1}}{( L_d\beta)^{\frac{d-1}{2}}}
                     \int_0^{\infty}\D\ts\,\ts^{\frac{d+1}{2}-1}e^{-\frac{\pi \rho_L^2}{\ts}}
                     \frac 12 \sideset{}{'}\sum_{(r,\ell)\in \mathbb{Z}^2}e^{-\frac{\pi t}{\tau_2}|r\tau+l|^2},\\
                   &=\frac{V_{d-1}}{( L_d\beta)^{\frac{d-1}{2}}}
                     \sideset{}{'}\sum_{(r,\ell)\in \mathbb{Z}^2}
                     \Big(\frac{\sqrt{\tau_2}\rho_L}{|r\tau+l|}
                     \Big)^{\frac{d+1}{2}}
                     K_{\frac{d+1}{2}}\Big(
                     2\pi \frac{\rho_L}{\sqrt \tau_2}|r\tau+l|\Big).
  \end{split}
\end{equation}
The latter result can be compactly expressed in terms of the recently defined
modular invariant massive deformations of Maass-Jacobi forms
\cite{Berg:2019jhh},
\begin{equation} \label{Maass Jacobi form}
  \mathcal{E}_{s,\mu}(0;\tau)=2\sideset{}{'}\sum_{(r,\ell)\in \mathbb{Z}^2}\left (\sqrt{\mu\tau_2}\over |r\tau+\ell|\right )^sK_s
  \left (2\pi\sqrt{\mu\over\tau_2}|r\tau+\ell|\right ),
\end{equation}
with the identification $\mu=\rho^2_L$ as
\begin{equation}
  \label{eq:25}
  \ln {Z}_{d,1}(g,m)=\frac{V_{d-1}}{2( L_d\beta)^{\frac{d-1}{2}}} \mathcal{E}_{\frac{d+1}{2},\rho_L^2}(0;\tau).
\end{equation}
From the canonical approach one gets,
\begin{multline}
  \label{eq:37e}
  \ln Z_{d,1}(\tau_1,\tau_2,L_d,m)=\frac{V_{d-1}}{L_d^{d-1}}
  \Big[2\tau_2
  \sum_{l\in\mathbb N^*}
  \Big(\frac{\rho_H}{l}
  \Big)^{\frac{d+1}{2}}
  K_{\frac{d+1}{2}}\Big(
  2\pi \rho_Hl\Big)+ \\
  2\tau_2^{-\frac{d-2}{2}}
  \sum_{l\in \mathbb{N}^*}\sum_{n\in
    \mathbb Z}\Big(\frac{\sqrt{n^2+\rho_H^2}}{l}\Big)^{\frac{d}{2}}
  K_{\frac{d}{2}}\Big(2\pi l \tau_2 \sqrt{n^2+\rho_H^2}\Big)e^{2\pi i l n \tau_1}
  \Big].
\end{multline}

\subsection{No small spatial dimensions: massive scalar black body}
\label{sec:no-small-spatial}

The massive scalar black body corresponds to the case where there is no small
spatial dimension, $q=0$, $\text{V}_{1}=\beta$, $\text{V}_0=1$,
$\rho_L=\frac{\beta m}{2\pi}$, $\rho_H=\frac{m}{2\pi}$, $b=\beta=b_D^{-1}$. The partition function
simplifies to
\begin{equation}
  \label{eq:17d}
  \begin{split}
  \ln {Z}_{d,0}(\beta,m)&=\frac{V_{d}}{\beta^d}
                   \int_0^{\infty}\D\ts\,\ts^{\frac{d+1}{2}-1}e^{-\frac{\pi \rho_L^2}{\ts}}\frac 12 \big(\theta(t)-1\big)\\
                 &=\frac{V_{d}}{\beta^d}\sum_{l\in \mathbb N^*}2\Big(\frac{\rho_L}{l}
                   \Big)^{\frac{d+1}{2}}
                   K_{\frac{d+1}{2}}(
                   2\pi \rho_Ll).
  \end{split}
\end{equation}
There is a scaling symmetry which follows from dimensional analysis: if the
scale is denoted by
$\lambda>0$,
\begin{equation}
  \label{eq:10}
  \ln {Z}_{d,0}(\lambda\beta,\lambda^{-1}m)=\lambda^{-d}\ln Z_d
  (\beta,m).
\end{equation}

The massless limit $\rho_L\to 0^+$ yields the scalar black body result
\begin{equation}
  \label{eq:9}
  \ln {Z}_{d,0}(\beta,0)=\frac{V_{d}}{\beta^d}\frac{\Gamma(\frac{d+1}{2})}{\pi^{\frac{d+1}{2}}}\zeta(d+1).
\end{equation}

\section{Bose-Einstein condensation} \label{sec:BEC}

In this section, we briefly discuss Bose-Einstein condensation from the
perspective of the exact expressions of the partition function of a complex
charged relativistic massive scalar field.

\subsection{Low temperature regime}
\label{sec:low-temp-volume}

The charge density is defined as
\begin{equation}
\label{eq:47b}
\delta_{d,q}=\frac{\langle Q \rangle}{V_{d-q}}=\frac{1}{{V_{d-q}}\beta}\frac{\partial}{\partial\nu}\ln Z_{d,q}(e,m,\nu).
\end{equation}
Note that the maximum value of $ |\nu| $ is $ m $. This can be most easily seen
from \eqref{eq:69} and \eqref{eq:7a} since there is a branch point at
$ |\nu|=m $ and the partition function ceases to be real for $ |\nu|>m $. The
fact that this is a special value can also be understood from \eqref{eq:56a} and
\eqref{eq:49} because the effective mass squared, $ m^2-\nu^2 $, becomes
negative.

Using \eqref{eq:37a}, it is explicitly given by
\begin{multline}
\label{eq:rho}
\delta_{d,q}=\frac{4b^{-\frac{d-q-1}{2}}}{( \text{V}_q)^{\frac{d-q}{q}}}
\Big[
\sum_{l\in \mathbb{N}^*}\sum_{n_\iota\in
	\mathbb Z^q}\Big (\frac{\sqrt{n_\iota\hat
		h^{\iota\kappa}n_\kappa+\rho_H^2}}{l}\Big )^{\frac{d+1-q}{2}}
l K_{\frac{d+1-q}{2}}\Big (2\pi l b \sqrt{n_\iota\hat
	h^{\iota\kappa}n_\kappa+\rho_H^2}\Big )\\e^{2\pi i l n_\iota N^\iota}\sinh{(l\beta\nu)}
\Big].
\end{multline}

In the low temperature regime, $ b\gg 1 $ and $ \beta m\gg 1 $. Also, using
$ n_\iota \hat h^{\iota \kappa}n_{\kappa} \geq 1,\; \forall n_{\iota}\neq (0,\cdots, 0)$, $ 2\pi l b\rho_H=l\beta m $, and the
asymptotic behavior of Bessel functions \eqref{eq:78}, the terms with
$n_{\iota}\neq (0,\cdots, 0)$ are suppressed by a factor $e^{-2\pi l b(\sqrt{1+\rho_H^2}-\rho_H)}$
with respect to those with $n_{\iota}=0$. Hence, to leading order
\begin{equation}
  \label{eq:densityleading}
  \delta_{d,q}\approx \frac{4b^{-\frac{d-q-1}{2}}}{( \text{V}_q)^{\frac{d-q}{q}}}
\sum_{l\in \mathbb{N}^*}\Big(\frac{\rho_H}{l}\Big)^{\frac{d+1-q}{2}}
l K_{\frac{d+1-q}{2}}(l\beta m)\sinh{(l\beta\nu)}.
\end{equation}
Note that, if there are only large dimensions ($q=0$), this is the exact result
since there are no terms with $n_{\iota}\neq (0,\cdots, 0)$,
\begin{equation}
  \label{eq:12}
  \delta_{d,0}= 4\beta^{-d} \big(\frac{\beta m}{2\pi}\big)^{\frac{d+1}{2}}
  \sum_{l\in \mathbb{N}^*}l^{-\frac{d-1}{2}}
  K_{\frac{d+1}{2}}(l\beta m)\sinh{(l\beta\nu)}.
\end{equation}

In the general case $q\neq 0$, up to terms of order
$\mathcal O\big ((\beta m)^{-1}\big )$, the leading order can be expanded
further as
\begin{equation}
  |\delta_{d,q}|\approx{2}\beta^{-(d-q)}
  \sum_{l\in \mathbb{N}^*}\Big(\frac{\beta m}{2\pi l}\Big)^{\frac{d-q}{2}}
e^{-l\beta (m-|\nu|)},
\end{equation}
where the sign of $\delta_{d,q}$ follows that of $\nu$. The critical density is
\begin{equation}
  |\delta^c_{d,q}|=	| \delta_{d,q}(|\nu|=m)|\approx {2}\Big(\frac{m}{2\pi \beta}\Big )^{\frac{d-q}{2}}
\sum_{l\in \mathbb{N}^*}\Big(\frac{1}{l}\Big)^{\frac{d-q}{2}}.
\end{equation}
The sum over $ l $ converges if and only if $ d-q> 2 $. There should thus be at
least $3$ large dimensions to achieve Bose-Einstein condensation, and
\begin{equation}
  |\delta^{c}_{d,q}|\approx{2}\Big(\frac{m}{2\pi \beta}\Big )^{\frac{d-q}{2}}\zeta({d-q\over 2}), \quad d-q>2.
\end{equation}
The critical temperature is
\begin{equation}
  T_c\approx{2\pi\over m}\Big ( { |\delta_{d,q}|\over 2}\zeta({d-q\over 2})\Big )^{2\over d-q}.
\end{equation}
The condensate is formed when the charge density is greater than the critical
charge density and the charge density of the ground state/condensate is
$ \delta^{G}_{d,q}= \delta_{d,q}- \delta^{c}_{d,q}$
\begin{equation}
\label{eq:charge density ground}
\begin{cases}T>T_c:   \delta^G_{d,q}=0,\\
  T<T_c: \frac{ \delta^G_{d,q}}{ \delta_{d,q}}=1-(\frac{T}{T_c})^{d-q\over 2}
\quad{\rm and }\; \nu=\pm m.
\end{cases}
\end{equation}

\subsection{High temperature regime}
\label{sec:high-temp-regime}

In the high temperature/large volume expansion, the charge density $\delta_{d,q}$ can
be found using \eqref{eq:37c} and the identity \eqref{eq:XXX} at $n=1$,
\begin{multline} \label{eq:rhohigh}
  \delta_{d,q}=\frac{1}{( \text{V}_{q}^D)^{\frac{d-q}{q}}}
	\Big[
	4b_D^d\sum_{l\in \mathbb{N}^*}l\sinh{(l\beta\nu)}\Big (\frac{\rho_H^D}{l b_D}\Big)^{\frac{d+1}{2}}
	K_{\frac{d+1}{2}}\Big (2\pi l \frac{\rho_H^D}{b_D}\Big )\\
	-b_D^{\frac{d}{2}} \sum_\pm \sideset{}{'}\sum_{m^\iota\in
		\mathbb Z^q}\sum_{n\in \mathbb
		Z}
	\pm \Big(\frac{\sqrt{(\frac{\rho_H^D}{b_D})^2+(n\pm i \frac{\beta\nu}{2\pi})^2}}{\sqrt{m^\iota\hat
			h^D_{\iota\kappa}m^\kappa}}\Big)^\frac{d}{2} m^{\iota}N_{\iota}^D
	\\K_{\frac{d}{2}}\Big (2\pi b_D
	\sqrt{(\frac{\rho_H^D}{b_D})^2+(n\pm i \frac{\beta\nu}{2\pi})^2}\sqrt{m^\iota\hat
		h^D_{\iota\kappa}m^\kappa}\Big)e^{2\pi im^\iota N_\iota^D n\mp m^\iota N_\iota^D\beta\nu}\\
		-i b_D^{\frac{d}{2}+1}\sum_\pm\sideset{}{'}\sum_{m^\iota\in
		\mathbb Z^q}\sum_{n\in \mathbb
		Z}
	\pm \Big(\frac{\sqrt{(\frac{\rho_H^D}{b_D})^2+(n\pm i \frac{\beta\nu}{2\pi})^2}}{\sqrt{m^\iota\hat
			h^D_{\iota\kappa}m^\kappa}}\Big)^{\frac{d}{2} -1}(n\pm i \frac{\beta\nu}{2\pi})
	\\K_{\frac{d}{2}-1}\Big (2\pi b_D
	\sqrt{(\frac{\rho_H^D}{b_D})^2+(n\pm i \frac{\beta\nu}{2\pi})^2}\sqrt{m^\iota\hat
		h^D_{\iota\kappa}m^\kappa}\Big)e^{2\pi im^\iota N_\iota^D n\mp m^\iota N_\iota^D\beta\nu}
	\Big].
\end{multline}
The high temperature/large volume regime may be defined through
\begin{equation}
  \label{eq:40a}
  b_D=\frac{( \text{V}_q)^{\frac{1}{q}}}{\beta (1+\mu_i\mu^i)^{\frac{q+1}{2q}}}\gg 1,\
  \rho_H^D=\frac{m( \text{V}_q)^{\frac{1}{q}}}{2\pi(1+\mu_i\mu^i)^{\frac{1}{2q}}}\gg 1,\
  {\rho_H^D\over b_D}=\frac{\beta m}{ 2\pi}\sqrt{1+\mu_i\mu^i} \ll 1.
\end{equation}
We will now argue that the first line of \eqref{eq:rhohigh} is the dominant one
in this regime. When using the asymptotic behavior of Bessel functions
\eqref{eq:78}, in both of the sums over $n$, the terms with $ n\neq0 $ are
exponentially suppressed as compared to the term with $n=0$. When these are
neglected, the arguments of the Bessel functions in the third and fifth line of
\eqref{eq:rhohigh} become
\begin{equation}
  \label{eq:62}
  2\pi \rho_H^D\sqrt{1-\frac{\nu^2}{m^2(1+\mu_i\mu^i)}}{\sqrt{m^\iota\hat
			h^D_{\iota\kappa}m^\kappa}}.
\end{equation}
If this argument is large, these terms are still exponentially suppressed. A
necessary condition for the argument to become small is that
$\frac{\nu^2}{m^2(1+\mu_i\mu^i)}$ is very close to unity, and that $|\nu|$ is close to
$m$ with all $\mu^{i}$ close to zero. For simplicity, let us assume in the
following that $\mu^{i}=0$, which implies $N_{\iota}^{D}=0$ so that the terms on the
second and third lines vanish. We then have
\begin{multline}
  \label{eq:68}
  \delta_{d,q}\approx \frac{1}{( \text{V}_{q})^{\frac{d-q}{q}}}
	\Big[4b_D^d\sum_{l\in \mathbb{N}^*}l\sinh{(l\beta\nu)}\Big (\frac{\rho_H^D}{l b_D}\Big)^{\frac{d+1}{2}}
	K_{\frac{d+1}{2}}\Big (2\pi l \frac{\rho_H^D}{b_D}\Big )\\
  + 2b_D^{2}(\frac{\beta\nu}{2\pi})\sideset{}{'}\sum_{m^\iota\in
		\mathbb Z^q}
	\Big(\frac{\rho_H^D\sqrt{1-\frac{\nu^2}{m^2}}}{\sqrt{m^\iota\hat
      h^D_{\iota\kappa}m^\kappa}}\Big)^{\frac{d}{2} -1}\\
	K_{\frac{d}{2}-1}\Big  (2\pi\rho_H^D
	\sqrt{1-\frac{\nu^2}{m^2}}\sqrt{m^\iota\hat
		h^D_{\iota\kappa}m^\kappa}\Big)\Big].
\end{multline}
Let us now assume that one may do a Taylor expansion around $0$ of each of the
terms in the series using \eqref{BesselK limit}, even if the arguments of the
Bessel functions end up being large for sufficiently large $l$ or $m^{\iota}$ when
$|\nu|<m$. The validity of this assumption is discussed in Appendix \ref{check}. One finds
\begin{multline}
  \label{eq:73}
  \delta_{d,q}\approx\frac{1}{( \text{V}_{q})^{\frac{d-q}{q}}}\Big[
  \frac{	2b_D^d\nu\beta}{\pi^{\frac{d+1}{2}}}\Gamma (\dfrac{d+1	}{2} ) \left(1-\mathcal O(m^2 \beta^2)\right)\sum_{l\in \mathbb{N}^*}\frac{1}{l^{d-1}}\\
+\frac{{b^{2}_D}\nu\beta}{2\pi^{\frac{d}{2}}}\Gamma (\frac{d}{2}-1)\left (1-\mathcal{O}(\beta^2(m^2-\nu^2))\right )\sideset{}{'}\sum_{m^\iota\in
		\mathbb Z^q}\frac{1}{(m^\iota h^D_{\iota \kappa}m^\kappa)^{\frac{d}{2}-1}}
	\Big].
\end{multline}
Note that the sum in the second term converges when $ d-q>2 $. There should thus
be at least $3$ large dimensions to achieve Bose-Einstein condensation. This is
in agreement with the analysis in the low-temperature expansion. Moreover, the
first line dominates for $ d>2 $. So, the critical density is
\begin{equation} \label{eq:critical density}
  \delta^{c}_{d,q}= \delta_{d,q}|_{\nu=\pm m}\approx\pm\frac{2m\text{V}_q}{\pi^{\frac{d+1}{2}}\beta^{d-1}}
	\Gamma (\dfrac{d+1	}{2})\zeta(d-1)~, \quad d-q>2.
\end{equation}
The critical temperature is
\begin{equation}
	T_c\approx\Big (
    \frac{\pi^{\frac{d+1}{2}}|\delta_{d,q}|}{2 m \text{V}_q\Gamma (\dfrac{d+1	}{2} )\zeta(d-1)}
  \Big )^{\frac{1}{d-1}}.
\end{equation}
The condensate is formed when the charge density is greater than the critical
charge density and the charge density of the ground state/condensate is
$  \delta^{G}_{d,q}= \delta_{d,q}- \delta^{c}_{d,q} $
\begin{equation}
	\label{eq:charge density ground high}
  \begin{cases}T>T_c:   \delta^G_{d,q}=0,\\
		T<T_c: \frac{ \delta^G_{d,q}}{ \delta_{d,q}}=1-(\frac{T}{T_c})^{d-1}
		\quad{\rm and }\; \nu=\pm m.
	\end{cases}
\end{equation}
For $3$ large dimensions, $d=3, q=0$, one recovers the well-known results.

\section{Conclusions}

We have studied the partition function of free massive scalar field theory on
locally flat manifolds of the form $\mathbb R ^{d-q}\times \mathbb T ^{q+1}$ with two
different types of chemical potentials turned on. The main goal of turning on
the first type of chemical potentials, $\mu^{i}$ for linear momentum in the small
directions, was to study the effect of mass on ${\rm SL}(q+1,\mathbb Z)$
invariance (and on modular invariance for $q=1$) by computing exact partition
functions. We found that, for $q=1$, the free energy $F=-\beta^{-1}\ln Z$ is
described by modular invariant massive deformations of Maass-Jacobi forms
\eqref{Maass Jacobi form} recently studied in \cite{Berg:2019jhh}. For $q>1$, we
obtain higher dimensional generalizations of these forms \eqref{eq:81} which are
${\rm SL}(q+1,\mathbb Z)$ invariant. Making use of this invariance, we obtained
the high and low-temperature expansions of the free energy.

By introducing a second chemical potential, $\nu$ for the charge of $U(1)$
rotations of a complex massive scalar, we generalized some of the known results
on relativistic Bose-Einstein condensation by Kapusta, Haber and Weldon
\cite{Kapusta:1981aa,Haber:1981ts,Haber:1981tr,Haber1981} to include arbitrary
number of small dimensions. For even $d$, these types of partition functions
have recently been shown to be related to conformal graphs
\cite{Karydas:2023ufs}. It would be interesting to explore the connection
between the recursion relations of \cite{Karydas:2023ufs} to those of
Haber-Weldon given in equation \eqref{eq:20e}. More generally, one might try to
understand whether there is a direct connection between these scalar field
partition functions and string theory scattering amplitudes, where these modular
invariant forms and their massive deformations also appear.

\section*{Acknowledgments}
\label{sec:acknowledgements}

\addcontentsline{toc}{section}{Acknowledgments}

The authors are grateful to A.~Kleinschmidt and A.~Petkou for useful
discussions. AA is supported by Austrian Science Fund (FWF), projects P 33789,
and P 36619. This work has been supported by the Fund for Scientific Research
(F.R.S.-FNRS) Belgium through a research fellowship for AA as well as
conventions FRFC PDR T.1025.14 and IISN 4.4503.15. AA was partially supported by
the Delta ITP consortium, a program of the NWO that is funded by the Dutch
Ministry of Education, Culture and Science (OCW). GB is grateful to the Mainz
Institute for Theoretical Physics (MITP) of the Cluster of Excellence PRISMA$^+$
(Project ID 39083149), for its hospitality and its support during the completion
of this work.

\appendix

\section{Definitions and useful formulae}
\label{sec:defin-usef-form}

\paragraph{Riemann theta function}

The Riemann theta function is defined by
\begin{equation}
  \label{F5}
  \vartheta_n(z|g)=\sum_{m^\alpha\in\mathbb{Z}^n}e^{\pi i m^\alpha g_{\alpha\beta} m^\beta+2\pi im^
    \alpha z_\alpha},\qquad \alpha,\beta=1,...,n,
\end{equation}
for $g=g_{1}+ig_{2}\in\mathbb{H}_n=\{g\in M_{n}(\mathbb{C})|g=g^T, g_2>0\}$, with
$M_{n}(\mathbb{C})$ complex $n\times n$ matrices and
$z\in\mathbb{C}^n$. Let also
\begin{equation}
  \label{F6}
  \theta_n(g)\equiv\vartheta_n(0|ig)=\sum_{m^\alpha\in\mathbb{Z}^n}
  e^{-\pi m^\alpha g_{\alpha\beta} m^\beta}.
\end{equation}
and
\begin{equation}
  \label{eq:8}
  \theta(t)=\sum_{m\in\mathbb{Z}}
  e^{-t \pi m^2}.
\end{equation}

\paragraph{Poisson resummation formula}

\begin{multline}
  \label{W11}
  \sum_{m^\alpha\in\mathbb{Z}^n}e^{-\pi
    (m^{\alpha}+v^{\alpha})A_{\alpha\beta}(m^{\beta}+v^{\beta})
    +2\pi i m^{\alpha}w_{\alpha}}
  \\=\frac{1}{\sqrt{\mathrm{det}\, A}}
  \sum_{n_\alpha\in\mathbb{Z}^n}e^{-\pi (n_{\alpha}+w_{\alpha})
    (A^{-1})^{\alpha\beta}(n_{\beta}+w_{\beta})
    -2\pi i (n_{\alpha}+w_{\alpha})v^{\alpha}}.
\end{multline}

\paragraph{Integrals over Schwinger time}

The following integrals are needed:
\begin{equation}
  \label{Gamma}
  \int_0^\infty d\ts \,\ts^{\nu-1}e^{-\ts \mu}={\Gamma(\nu)\over \mu^\nu},
\end{equation}
for [${\rm Re}\,(\mu)>0$, ${\rm Re}\,(\nu)>0$],
(see \cite{gradshteyn2007}, 3.381.4),
and also
\begin{equation}
  \label{BesselK}
  \int_0^\infty d\ts\, \ts^{s-1}e^{-{\mu\over \ts}-\nu\ts}=2({\mu\over \nu})^{\frac s2}{K}_{s}(2 \sqrt{\mu \nu}),\quad
\end{equation}
for [${\rm Re}\,\mu>0$, ${\rm Re}\,\nu>0$] (see \cite{gradshteyn2007}, 3.471.9).

Here ${K_{s}(z)}=K_{-s}(z)$ is the modified Bessel function of the second type.
It also follows that for small $x$,
\begin{equation}
  \label{BesselK limit}
x^s K_s(x)=2^{s-1}\Gamma(s)\left (1-\frac{x^2}{4(s-1)}+\mathcal O (x^4)\right ),\quad s>0.
\end{equation}
Furthermore,
\begin{equation}
  \label{eq:XXX}
  \big(\frac{d}{zdz}\big)^n	(z^s K_{s}(z))=(-1)^n z^{s-n} K_{s-1}(z).
\end{equation}
for [$n\in \mathbb N$] (see \cite{gradshteyn2007}, 8.486.14).

\paragraph{Integrals in the functional approach}

\begin{equation}
  \label{eq:64}
  \int^\infty_0dx\,\frac{\sinh^\mu x}{\cosh^\nu x}=\frac 12 B(\frac{\mu+1}{2},\frac{\nu-\mu}{2})=
  \frac 12 \frac{\Gamma(\frac{\mu+1}{2})
    \Gamma(\frac{\nu-\mu}{2})}{\Gamma(\frac{\nu+1}{2})},
\end{equation}
for [${\rm Re}\,\mu>-1$, ${\rm Re}\,(\mu-\nu)<0$] (see \cite{gradshteyn2007},
3.512.2).

\paragraph{Integrals in the canonical approach}

\begin{equation}
  \label{eq:289a}
  \int^\infty_udx\, x(x^2-u^2)^{\nu-1}e^{-x}=2^{\nu-\frac 12}
  \frac{\Gamma(\nu)}{\sqrt \pi} u^{\nu+\frac 12}K_{\nu+\frac{1}{2}}(u),
\end{equation}
for [$0< u\in \mathbb R$] (see \cite{gradshteyn2007}, 3.389.4).

\paragraph{Asymptotic expansion}

\begin{equation}
  \label{eq:78}
  K_\nu(z)=\sqrt{\frac{\pi}{2z}}e^{-z}(1+\mathcal O(z^{-1})),
\end{equation}
or, more precisely,
\begin{equation}
  \label{eq:87}
  K_\nu(z)=\sqrt{\frac{\pi}{2z}}e^{-z}\sum_{l=0}^\infty\frac{1}{(2z)^l}\frac{\Gamma(\nu+l+\frac 12)}{l!\Gamma(\nu-l+\frac 12)},
\end{equation}
for [$|z|\to \infty$] (see e.g.~formula (4.12) of \cite{Actor:1986zf} and
\cite{gradshteyn2007}, 8.451.6 for more details). For $\nu=n+\frac 12$, this
series cuts at $l=n$, and there is
an exact expansion in terms of polynomials,
\begin{equation}
  \label{eq:88}
  K_{n+\frac 12}(z)=\sqrt{\frac{\pi}{2z}}e^{-z}\sum_{l=0}^n\frac{1}{(2z)^l}\frac{(n+l)!}{l!(n-l)!},
\end{equation}
(see \cite{gradshteyn2007}, 3.468).

\paragraph{Zeta function}

For ${\rm Re}(s)>1$, the zeta function is defined by
\begin{equation}
  \label{eq:89}
  \zeta(s)=\sum_{n=1}^\infty\frac{1}{n^s}=\frac{1}{\Gamma(s)}\int^\infty_0dx\frac{x^{s-1}}{e^x-1}.
\end{equation}

\paragraph{Polylogarithms}

For $|z|<1$, polylogarithms $ { \rm Li}_s(z) $ are defined as
\begin{equation}
  \label{eq:31}
  {\rm Li}_s(z)=\sum_{n=1}^\infty\frac{z^n}{n^s}.
\end{equation}
Alternatively, they may also be defined from the Bose-Einstein integral
expression,
\begin{equation}
	\label{eq:22b}
	{\rm Li}_s(z)=\frac{1}{\Gamma(s)}\int^\infty_0dt\frac{t^{s-1}}{e^tz^{-1}-1}
	=-\frac{1}{\Gamma(s)}\int^1_0dt \frac{(-\ln t)^{s-1}}{t-z^{-1}},
\end{equation}
and satisfy
\begin{equation}
	\begin{split}
		\label{eq:27b}
		{\rm Li}_{s+1}(z)=\int^z_0dt\ \frac{{\rm Li}_s(t)}{t}\iff {\rm Li}_{s+1}(e^{-y})=\int^\infty_ydy'\ {\rm Li}_s(e^{-y'}),\\
		{\rm Li}_1(z)=-\ln (1-z),\quad
		z\frac{\partial {\rm Li}_s(z)}{\partial z}= {\rm Li}_{s-1}(z),\quad {\rm Li}_s(1)=\zeta(s).
	\end{split}
\end{equation}

Taking $ z=e^y $, the expansion around $ y=0 $ is given for integer $n$ by
\begin{equation}
	\label{eq:38 b}
	{\rm Li}_n(e^y)=\frac{y^{n-1}}{(n-1)!}[S_{n-1}-\ln(-y)]+\sum_{k=0,k\neq n-1}\frac{\zeta(n-k)}{k!}y^k.
\end{equation}
with  $-y^{n-1}\ln(-y)\to 0$ for $y\to 0$ except when $n=1$, and
\begin{equation}\label{key}
	S_k=\gamma+\psi(k+1)=\sum_{p=1}^k\frac 1p, \quad S_0=1.
\end{equation}
$\psi(z)={d\over dz}\ln \Gamma(z)$ is the digamma function and
$ \gamma=-\psi(1) $ is the Euler-Mascheroni constant.

\section{Computations in the wordline formalism}
\label{sec:comp-wordl-form}

The heat kernel \eqref{eq:20} may be directly worked out from simple building
blocks as follows : for a particle on $\mathbb R$,
\begin{equation}
  \label{eq:42}
  \langle  x'|e^{-t p^2}|x \rangle=\frac{1}{\sqrt{4\pi t}}e^{-\frac{(x'-x)^2}{4t}},
\end{equation}
while for a particle on a circle of radius $R$,
\begin{equation}
  \label{eq:45}
  \langle  x'|e^{-t p^2}|x \rangle={1\over R}\sum_{n=-\infty}^{\infty}e^{-\ts({2\pi n\over R})^2+i2\pi n{x-x'\over R}}\,
  = \, {1\over \sqrt{4\pi\ts}}\sum_{l=-\infty}^{\infty}e^{-{R^2[{x-x'\over R}+l]^2\over 4\ts}},
\end{equation}
and finally
\begin{equation}
  \label{eq:47}
  \langle A| e^{t\, \epsilon} |B\rangle=\cos{t}\,\delta_{AB}+\sin{t}\,\epsilon_{AB}.
\end{equation}
It follows that
\begin{multline}
\label{eq:54}
K_{\mathbb{R}^p\times\mathbb{T}^{q+1}}(x'^I,y'^\alpha,x^I,y^\alpha;\ts)_{AB}=
\frac{e^{-\ts(m^2-\nu^2)}}{(4\pi \ts)^{\frac{p}{2}} \text{V}_{q+1}}
e^{-\frac{1}{4\ts}(x'_I-x_I)(x'^I-x^I)}\\\frac 12 \sum_{n_\alpha\in\mathbb Z^{q+1}}
\Big[\Big(e^{-\ts[(2\pi)^2n_\alpha g^{\alpha\beta}n_\beta
-i4\pi\nu e\indices{_{d+1}^\alpha}n_\alpha]+2\pi i n_\alpha(
y'^\alpha-y^\alpha)}+(\nu\to -\nu)\Big)\delta_{AB}\\
-i\Big(e^{-\ts[(2\pi)^2n_\alpha g^{\alpha\beta}n_\beta
-i4\pi\nu e\indices{_{d+1}^\alpha}n_\alpha]+2\pi i n_\alpha(
y'^\alpha-y^\alpha)}-(\nu\to -\nu)\Big)\epsilon_{AB}
\Big].
\end{multline}
Taking the trace and redefining the Schwinger proper time
$t\to \frac{t}{4\pi}(\text{V}_{q+1})^{\frac{2}{q+1}}$ so as to factorize the volume
yields,
\begin{multline}
  \label{eq:15a}
  \ln Z_{d,q}(e,m,\nu)
  =\frac{V_{d-q}}{2( \text{V}_{q+1})^{\frac{d-q}{q+1}}}
  \int_0^{\infty}\D\ts\,\ts^{\frac{q-d}{2}-1}\\\sum_\pm\sum_{n_\alpha\in\mathbb
    Z^{q+1}}
  e^{-\pi \ts [\rho_L^2-\nu^2_L+n_\alpha \hat g^{\alpha\beta}n_\beta\mp 2i \nu_L{\hat
    e}\indices{_{d+1}^\alpha}n_\alpha]},
\end{multline}
with $\rho_L,\nu_L$ defined in \eqref{eq:59}.
Defining furthermore
\begin{equation}
  v_\pm^\alpha= \pm \nu_L { \hat e}\indices{_{d+1}^\alpha},\quad w^\pm_\alpha=\mp i\nu_{L} { \hat e}\indices{^{d+1}_\alpha},
  \label{eq:60}
\end{equation}
the partition function \eqref{eq:15a} admits a compact integral representation in terms
of the Riemann theta function $\vartheta_{q+1}(z|g)$ defined in \eqref{F5},
\begin{equation}
  \label{eq:15b}
  \ln Z_{d,q}(e,m,\nu)
  =\frac{V_{d-q}}{2( \text{V}_{q+1})^{\frac{d-q}{q+1}}}
  \int_0^{\infty}\D\ts\,\ts^{\frac{q-d}{2}-1}e^{-\pi t(\rho_L^2-\nu^2_L)}\sum_\pm\vartheta_{q+1}(t v_{\pm}|ti \hat g^{-1}).
\end{equation}
The argument in the exponential of \eqref{eq:15a} may
be re-written as
\begin{equation}
  \label{eq:61}
  \rho_L^2-\nu^2_L+n_\alpha \hat g^{\alpha\beta}n_\beta\mp 2i \nu_L{\hat
    e}\indices{_{d+1}^\alpha}n_\alpha=\rho^2_{L}+(n_\alpha+w^\pm_\alpha) \hat g^{\alpha\beta}(n_\beta+w^\pm_\beta),
\end{equation}
where the $w^\pm_\alpha$ are defined in \eqref{eq:60}. This may be used in order to
apply the Poisson resummation formula \eqref{W11}, which gives,
\begin{equation}
  \label{eq:59a}
  \ln Z_{d,q}(e,m,\nu)
  =\frac{V_{d-q}}{2( \text{V}_{q+1})^{\frac{d-q}{q+1}}}
  \int_0^{\infty}\D\ts\,\ts^{-\frac{d+1}{2}-1}\sum_\pm
  \sum_{m^\alpha\in\mathbb
    Z^{q+1}}
  e^{-\pi \ts \rho_L^2-\frac{\pi}{\ts} m^\alpha \hat g_{\alpha\beta}m^\beta+2\pi i  m^{\alpha}w^\pm_\alpha}.
\end{equation}

Note that, when using \eqref{Gamma}, the $m^{\alpha}=0$ zero mode gives rise to the contribution,
\begin{equation}
  \label{eq:67}
  \frac{V_{d-q}}{(\text{V}_{q+1})^{\frac{d-q}{q+1}}}
  \pi^{\frac{d+1}{2}}\Gamma(-\frac{d+1}{2})\rho_L^{d+1}=
  V_{d-q}\text{V}_{q+1}m^{d+1}\frac{\Gamma(-\frac{d+1}{2})}{2^{d+1}\pi^{\frac{d+1}{2}}},
\end{equation}
to the partition function, which diverges for odd $d$. This term is
proportional to the total volume of the Euclidean spacetime box but independent
of chemical potentials. It is thus manifestly $ {\rm SL}(q+1,\mathbb{Z}) $
invariant. Therefore, this term can be removed without affecting this
invariance. Furthermore, the removal can be interpreted as shifting the zero of
the energy, i.e., by adding a suitable constant to the Hamiltonian density. Even
though this contribution is finite for even $d$, we choose to remove it both for
odd and even $d$ (see e.g.~\cite{Ambjorn:1981xw}).

After this subtraction, the final result is expressed in terms of the Riemann
theta function in \eqref{eq:59b}.

\section{Computations in the functional approach}
\label{sec:comp-funct-appr}

The eigenvalue matrix defined in \eqref{eq:51} is explicitly given by
\begin{equation}
  \label{eq:58}
  (\lambda^2_{n_I,n_\alpha})_{AB}=(\omega_S^2+\omega_a\omega^a)\delta_{AB}
  +2\nu \omega_{d+1}\epsilon_{AB},
\end{equation}
where
\begin{equation}
  \label{eq:55}
  \omega_a=2\pi e\indices{_a^\alpha}n_\alpha,\quad \omega_S^2=k_Ik^I+m^2-\nu^2.
\end{equation}
The associated determinant in the two-dimensional internal space is
\begin{multline}
  \label{eq:56}
  {\rm det}\, (\lambda^2_{n_I,n_\alpha})=  (\omega_S^2+\omega_a\omega^a)^2
  +4(\nu \omega_{d+1})^2\\=(\omega_S^2+\omega_a\omega^a-2i\nu\omega_{d+1})(\omega_S^2+\omega_b\omega^b+2i\nu\omega_{d+1}).
\end{multline}
When taking into account that the first $p$ dimensions become large, the sum
over eigenvalues $\sum_{n_I\in \mathbb Z^p}$ is replaced by an integral
$\frac{V_p}{(2\pi)^p}\int d^pk$. The latter is done in hyperspherical
coordinates, where integration over the angles gives the volume of the
hypersphere, with only an integral over the radius to be preformed. If $\lambda$
is a parameter of dimension inverse length, the associated zeta function is then
\begin{multline}
  \label{eq:63}
  \zeta_{\mathcal D}(s)=\frac{(2\pi\lambda^2)^sV_{p}}
  {(2\pi)^{p}}\frac{2\pi^{\frac{p}{2}}}{\Gamma(\frac{p}{2})}
  \sum_{n^\alpha\in \mathbb Z^{q+1}}\int_0^\infty d k k^{p-1}\\
  \Big[(\omega_S^2+\omega_a\omega^a-2i\nu\omega_{d+1})^{-s}+
    (\omega_S^2+\omega_a\omega^a+2i\nu\omega_{d+1})^{-s} \Big],
\end{multline}
with $\omega_S^2=k^2+m^2-\nu^2$. If ${\rm Re}(s)>\frac {p}{2}$, the integral is
convergent and can be done through the change of variables
$k=\sqrt{m^2-\nu^2+\omega_a\omega^a\mp 2i\nu\omega_{d+1}}\sinh x$. After using
\eqref{eq:64}, one finds
\begin{equation}
  \label{eq:65}
  \zeta_{\mathcal D}(s)=\frac{(2\pi\lambda^2)^sV_{d-q}}
  {(2\pi)^{p}}\frac{\pi^{\frac{p}{2}}\Gamma(s-\frac{p}{2})}{\Gamma(s)}
  \sum_{\pm}\sum_{n^\alpha\in \mathbb Z^{q+1}}(m^2-\nu^2+\omega_a\omega^a\mp
  2i\nu\omega_{d+1})^{\frac{p}{2}-s}.
\end{equation}
or, after factorizing $[2\pi (\text{V}_{q+1})^{-\frac{1}{q+1}}]^{p-2s}$,
\begin{equation}
  \label{eq:79}
  \zeta_{\mathcal D}(s)=\frac{(2\pi\lambda^2)^sV_{p}\pi^{\frac{p}{2}-2s}\Gamma(s-\frac{p}{2})}
  {2^{2s}(\text{V}_{q+1})^{\frac{p-2s}{q+1}}\Gamma(s)}
  \sum_{\pm}\sum_{n^\alpha\in \mathbb Z^{q+1}}(\rho_L^2-\nu_L^2+n_\alpha\hat g^{\alpha\beta}n_\beta\mp
  2i\nu_L{\hat e}\indices{_{d+1}^\alpha}n_\alpha)^{\frac{p}{2}-s}.
\end{equation}
In these terms, when taking into account that $\Gamma(s)^{-1}=s+\mathcal O(s^2)$,
the partition function \eqref{eq:71} can be formally written as
\begin{equation}
  \label{eq:80}
  \ln Z_{d,q}(e,m,\nu)=\frac{V_{d-q}\pi^{\frac{d-q}{2}}\Gamma(\frac{q-d}{2})}{2(\text{V}_{q+1})^{\frac{d-q}{q+1}}}
  \sum_{\pm}\sum_{n^\alpha\in \mathbb Z^{q+1}}(\rho_L^2-\nu_L^2+n_\alpha\hat g^{\alpha\beta}n_\beta\mp
  2i\nu_L{\hat e}\indices{_{d+1}^\alpha}n_\alpha)^{\frac{d-q}{2}}.
\end{equation}
This formal expression, which has to be analytically continued to the physical
values of $d-q$, can be also be obtained by performing the integration over
Schwinger proper time in \eqref{eq:15a} using \eqref{Gamma}. The analytic
continuation may be obtained by using \eqref{BesselK} in order to perform the
integration over Schwinger proper time in \eqref{eq:59a}, which gives the
representation \eqref{eq:81}, after the $m^\alpha=0$ mode has been excluded.

\section{Computations in the canonical approach}
\label{sec:compcan}

With the expression of the operators in \eqref{FF9}, direct evaluation of the
trace in \eqref{eq:3} gives
\begin{multline}
  \label{FF10} Z_{d,q}(\beta,\mu^j,h,m,\nu)
  =e^{-2 \beta E^{d,q}_0(h,m)}\prod_{(n_I,n_{\iota})\in\mathbb{Z}^d,\pm}
  \sum_{N_{k_i}}e^{(-\beta\omega_{k_i}+2\pi
    i\beta\mu^j\theta_j{}^{\iota}
    n_{\iota}\pm \beta\nu)N_{k_i}}\\=\prod_{(n_I,n_{\iota})\in\mathbb{Z}^d,\pm}\frac{e^{-2\beta
      E^{d,q}_0(h,m)}}{1-e^{-\beta\omega_{k_i}+2\pi
      i\alpha^j\theta_j{}^{\iota} n_{\iota}\pm\beta\nu}}.
\end{multline}
Taking the logarithm and turning the sums over $n_{I}$ into integrals gives
\begin{multline}
  \label{eq:69}
  \ln Z_{d,q}(\beta,\mu^j,\theta\indices{_i^\iota},m,\nu)=-2\beta E_0^{d,q}(h,m)\\-
  \frac{V_{d-q}}{2^{d-q-1}\pi^{\frac{d-q}{2}}\Gamma(\frac{d-q}{2})}
  \sum_\pm\sum_{n_{\iota}\in\mathbb{Z}^q}I_{d-q}(\beta,N^\iota,m,n^\iota,\pm\nu),
\end{multline}
where
\begin{equation}
  \label{eq:7a}
  I_{d-q}(\beta,N^\iota,m,n^\iota,\pm\nu)=\int_0^{\infty}\D k\,
  k^{d-q-1}\ln[1-e^{\pm\beta\nu}e^{-\beta\omega_{k_i}+2\pi
    iN^{\iota} n_{\iota}}].
\end{equation}
Introducing the variable $z$ as
\begin{equation}
  \label{eq:8b}
  \begin{split}
    & z=\beta\sqrt{k^2+(2\pi)^2n_{\iota}h^{\iota\kappa}n_{\kappa}+m^2},\\ &
  k^{d-q-1}dk=\frac{1}{\beta^{d-q}}
     \big[z^2-(2\pi\beta)^2n_{\iota}
      h^{\iota\kappa}n_{\kappa}-(m\beta)^2\big]^{\frac{d-q-2}{2}} z dz,
  \end{split}
\end{equation}
gives
\begin{multline}
  \label{eq:9a}
  I_{d-q}(\beta,N^\iota,m,n^\iota,\pm\nu)=\frac{1}{\beta^{d-q}}
  \int^\infty_{\beta\sqrt{(2\pi)^2n_{\iota}h^{\iota\kappa}n_{\kappa}+m^2}} dz z\\
  \big[z^2-(2\pi\beta)^2n_{\iota}h^{\iota\kappa}n_{\kappa}-(m\beta)^2\big]^{\frac{d-q-2}{2}}
  \ln \big[1-e^{\pm\beta\nu}e^{-z+2\pi i N^\iota n_\iota}\big].
\end{multline}
Expanding the logarithm, $\ln (1-x)=-\sum_{l\in \mathbb N^*}\frac{x^l}{l}$, and
changing the integration variable to $y=lz$ yields
\begin{multline}\label{eq:10a}
  I_{d-q}(\beta,N^\iota,m,n^\iota,\pm\nu)=-\frac{1}{\beta^{d-q}}\sum_{l\in\mathbb{N}^*}
  \frac{e^{2\pi i l N^\iota n_\iota}e^{\pm l\beta\nu}}{l^{d-q+1}}\\
  \int^\infty_{l\beta\sqrt{(2\pi)^2n_{\iota}h^{\iota\kappa}n_{\kappa}+m^2}} dy y
  \big[y^2-(2\pi\beta l)^2n_{\iota}h^{\iota\kappa}n_{\kappa}-(m\beta
  l)^2\big]^{\frac{d-q-2}{2}} e^{-y}.
\end{multline}
When using the integral \eqref{eq:289a}, we get for $q<d$ that
\begin{multline}
  \label{eq:11b}
  I_{d-q}(\beta,N^\iota,m,n^\iota,\pm\nu)=-
  \frac{2^{d-q}\pi^{\frac{d-q}{2}}\Gamma(\frac{d-q}{2})}{\beta^{\frac{d-q-1}{2}}}\\
  \sum_{l\in\mathbb{N}}
  \Big(\frac{\sqrt{n_{\iota}h^{\iota\kappa}n_{\kappa}
        +(\frac{m}{2\pi})^2}}{l}\Big)^{\frac{d-q+1}{2}}K_{\frac{d-q+1}{2}}
  \Big(2\pi l\beta\sqrt{n_{\iota}h^{\iota\kappa}n_{\kappa}+(\frac{m}{2\pi})^2}\Big)e^{\pm l\beta\nu}e^{2\pi i l
    N^\iota n_\iota},
\end{multline}
With the definitions in \eqref{eq:224a}, \eqref{eq:66}, \eqref{eq:13}, it follows
that
\begin{multline}
  \label{eq:12d}
  \ln Z_{d,q}(\beta,\mu^j,\theta\indices{_i^\iota},m,\nu)=-2\beta E^{d,q}_0(h,m)+
  \frac{4V_{d-q}}{( \text{V}_q)^{\frac{d-q}{q}}b^{\frac{d-q-1}{2}}}\sum_{n_{\iota}\in\mathbb{Z}^q}
  \sum_{l\in\mathbb{N}^*}\\
  \Big(\frac{{\sqrt{n_{\iota}\hat h^{\iota\kappa}n_{\kappa}+\rho_H^2}}}{l}\Big)^{\frac{d+1-q}{2}}
  K_{\frac{d-q+1}{2}}
  \Big(2\pi l b\sqrt{n_{\iota}\hat h^{\iota\kappa}n_{\kappa}
      +\rho_H^2}\Big)\cosh{(l\beta\nu)}e^{2\pi i l N^\iota n_\iota}.
\end{multline}

In order to get the Casimir energy, one may start from the expression of the
partition function obtained in the worldline approach in \eqref{eq:59a} using
the ADM parametrization of the unimodular metric,
\begin{equation}
\label{eq:234}
\hat g^{\mathrm{ADM}}_{\alpha\beta}
=b^{-\frac{2}{q+1}}\Big(\begin{matrix}\hat h_{\iota \kappa} &
\hat h_{\kappa\lambda}N^{\lambda}\\\hat h_{\iota \lambda}N^{\lambda}
& b^2+\hat h_{\lambda \xi}N^{\lambda}N^{\xi}
\end{matrix}\Big),
\end{equation}
\begin{equation}
\label{eq:259}
\hat g^{\alpha\beta}_{\mathrm{ADM}}=b^{-\frac{2q}{q+1}}
\Big(\begin{matrix}b^{2}\hat{h}^{\iota\kappa}+N^{\iota}N^{\kappa} & -
N^{\kappa}\\-
N^{\iota} &1
\end{matrix}\Big).
\end{equation}
One then splits the sum over $m^\alpha$ into a piece with $m^{d+1}=0$ which
gives
\begin{equation}
\label{eq:19a}
\frac{V_{d-q}}{( \text{V}_{q+1})^{\frac{d-q}{q+1}}}\sum_{m^\iota\in \mathbb Z^q}
\int_0^{\infty}\D t\,t^{\frac{d+1}{2}-1}e^{-\frac{\pi \rho_L^2}{t}}
e^{-\pi t b^{-\frac{2}{q+1}}m^\iota \hat h_{\iota\kappa} m^\kappa}
\end{equation}
After the change of variables $t=b^{\frac{2}{q+1}}t'$, and renaming $t'$ to
$t$, this gives
\begin{multline}
\label{eq:24}
\frac{V_{d-q}}{( \text{V}_q)^{\frac{d-q}{q}}}b\sum_{m^\iota\in \mathbb Z^q}
\int_0^{\infty}\D t\,t^{\frac{d+1}{2}-1}e^{-\frac{\pi
	\rho^2_H}{t}}\theta_q(t\hat h)=
\frac{V_{d-q}}{( \text{V}_q)^{\frac{d-q}{q}}}b\Big[
\Gamma(-\frac{d+1}{2})(\pi \rho^2_H)^{\frac{d+1}{2}}\\
+2\sideset{}{'}\sum_{m^\iota\in \mathbb Z^{q}}
\Big(\frac{\rho_H}{\sqrt{m^\iota\hat h_{\iota\kappa}m^\kappa}}
\Big)^{\frac{d+1}{2}}
K_{\frac{d+1}{2}}\Big(
2\pi \rho_H\sqrt{m^\iota\hat h_{\iota\kappa}m^\kappa}\Big)
\Big],
\end{multline}
where the first term on the right hand side is equal to the one discussed in
\eqref{eq:67}. As already discussed, it is subtracted both in even and in odd
dimensions. The second term corresponds to $-2\beta E_0^{d,q}$, the contribution
to the partition function associated to twice the renormalized Casimir energy
for a single massive scalar field, which is thus given by \eqref{eq:74}.

When using this result in \eqref{eq:12d}, one finds for the partition function
the expression given in \eqref{eq:37a}.

\section{Connection between worldline and canonical approach}
\label{sec:conn-betw-worldl}

In order to complete the connection between the results obtained in the
worldline and the canonical approaches, one considers the terms in
\eqref{eq:59a} with $m^{d+1}\neq 0$. They are given by
\begin{multline}
\label{eq:26}
\frac{V_{d-q}}{( \text{V}_q)^{\frac{d-q}{q}}}b^{-\frac{d-q}{q+1}}
\int_0^{\infty}\D t\,t^{\frac{d+1}{2}-1}e^{-\frac{\pi
	\rho_L^2}{t}}\\\sideset{}{'}\sum_{m^{d+1}\in \mathbb Z}\sum_{m^\iota\in \mathbb
Z^q} e^{-\pi t b^{-\frac{2}{q+1}}[(m^\iota+N^\iota m^{d+1})
\hat h_{\iota\kappa}(m^\kappa+N^\kappa m^{d+1})+b^2(m^{d+1})^2]}\cosh{(m^{d+1}\beta\nu)}.
\end{multline}
After a Poisson resummation \eqref{W11} over the spatial indices one finds
\begin{multline}
\frac{V_{d-q}}{(\text{V}_q)^{\frac{d-q}{q}}}b^{\frac{2q-d}{q+1}}
\int_0^{\infty}\D t\,t^{\frac{d+1-q}{2}-1}\\\sideset{}{'}\sum_{m^{d+1}\in
\mathbb Z}
\sum_{n_\iota\in
\mathbb Z^q} e^{-\frac{\pi}{t} b^{\frac{2}{q+1}}(n_\iota\hat
h^{\iota\kappa}n_\kappa+\rho_H^2)-2\pi i n_\iota N^\iota m^{d+1}-\pi t
b^{-\frac{2}{q+1}}b^2(m^{d+1})^2}\cosh{(m^{d+1}\beta\nu)}.\label{eq:27}
\end{multline}
When changing variables $t=t'\pi b^{\frac{2}{q+1}}(n_\iota\hat
h^{\iota\kappa}n_\kappa+\rho_H^2)$ and using \eqref{BesselK}, one finds
\begin{multline}
\label{eq:28}
\frac{4V_{d-q}}{(\text{V}_q)^{\frac{d-q}{q}}b^{\frac{d-q-1}{2}}}
\sum_{l\in \mathbb{N}^*}\sum_{n_\iota\in \mathbb
  Z^q}\\\Big(\frac{\sqrt{n_\iota\hat h^{\iota\kappa}n_\kappa+\rho_H^2}}{l}\Big)^{\frac{d+1-q}{2}}
K_{\frac{d+1-q}{2}}\Big(2\pi b l \sqrt{n_\iota\hat
h^{\iota\kappa}n_\kappa+\rho_H^2}\Big)e^{2\pi i n_\iota N^\iota
l}\cosh{(l\beta\nu)},
\end{multline}
which agree with the remaining terms of \eqref{eq:12d}.

\section{Computations for the high temperature expansion}
\label{sec:comp-high-temp}

One starts by rewriting \eqref{eq:59a} written as,
\begin{multline}
\ln Z_{d,q}(e,m,\nu)\\
=\frac{V_{d-q}}{( \text{V}^D_{q})^{\frac{d-q}{q}}}
b_D^{\frac{d-q}{q+1}}\sum_{m^\alpha\in\mathbb
Z^{q+1}}\cosh{(m^{d+1}\nu\beta)}\int_0^{\infty}\D\ts\,\ts^{-\frac{d+1}{2}-1}
e^{-\pi \ts \rho_L^2-\frac{\pi}{\ts} m^\alpha \hat g^{\rm ADM}_{\alpha\beta}m^\beta}.
\end{multline}
One now uses the dual ADM parametrization of the unimodular metric,
\begin{align}
\label{eq:245.2}
&\hat g^{\mathrm{ADM}}_{\alpha\beta}=(b_D)^{\frac{-2q}{q+1}}\Big(\begin{matrix}b_D^2\hat{h}^D_{\iota \kappa}
+N^D_{\iota}N^D_{\kappa}& -N_{\kappa}^D
\\-N^D_{\iota} & 1
\end{matrix}\Big),\\
\label{eq:245.3}
&\hat g^{\alpha\beta}_{\mathrm{ADM}}
=(b_D)^{-\frac{2}{q+1}}\Big(\begin{matrix}\hat h^{\iota\kappa}_D & N^{\kappa}_D
\\N^{\iota}_D & b_D^2+\hat h^D_{\iota \kappa}N^{\iota}_DN^{\kappa}_D
\end{matrix}\Big),
\end{align}
and $\rho_L=\rho_H^Db_D^{-\frac{1}{q+1}}$ to write the argument of the exponential as
\begin{equation}
\label{eq:75}
-\pi \ts
b_D^{-\frac{2}{q+1}}(\rho_H^D)^2
-\frac{\pi}{\ts}b_D^{-\frac{2q}{q+1}}[m^\iota(b^2_D\hat
h^D_{\iota\kappa}+N_\iota^DN_\kappa^D)m^\kappa-2m^\iota m^{d+1} N_{\iota}^D + (m^{d+1})^2]
\end{equation}

One first splits the sum into a piece with $m^\iota=0$, and treats
separately the terms with $m^{d+1}=0$ and $m^{d+1}\neq
0$. Using \eqref{Gamma} and \eqref{BesselK}, this gives as
contribution to $\ln Z_{d,q}(e,m,\nu)$ the terms
\begin{multline}
  \label{eq:6}
  \frac{V_{d-q}}{( \text{V}^D_{q})^{\frac{d-q}{q}}}\Big[b_D^{-1}\pi^{\frac{d+1}{2}}(\rho_H^D)^{d+1}
  \Gamma\Big (-\frac{d+1}{2}\Big )
\\ +4b_D^d\sum_{l\in \mathbb{N}^*}\cosh{(l\nu\beta)}\Big (\frac{\rho_H^D}{l b_D}\Big )^{\frac{d+1}{2}}
  K_{\frac{d+1}{2}}\Big (2\pi l \frac{\rho_H^D}{b_D}\Big )\Big].
\end{multline}
The first term is the same as discussed in \eqref{eq:67} and the first term on
the right hand side of \eqref{eq:24} and is subtracted.

For the remaining sum with $m^\iota\neq 0$,
one performs a Poisson resummation \eqref{W11} on $m^{d+1}$, with
$A_{d+1d+1}=\ts^{-1}b_D^{-\frac{2q}{q+1}}$, $y^{d+1}=0$ and $z^{d+1}_\pm
=-i\ts^{-1}b_D^{-\frac{2q}{q+1}}\pm i \frac{\nu\beta}{2\pi}$ %
which gives, together with the remaining previous terms, the result
\eqref{eq:37c}. Note that, when separating the terms with $n=0$ from the others,
the remaining contributions can also be written as
\begin{multline}
\label{eq:76}
\frac{V_{d-q}}{( \text{V}^D_{q})^{\frac{d-q}{q}}}\sideset{}{'}\sum_{m^\iota\in
\mathbb Z^q}2\cosh{(m^\iota N_\iota^D\nu \beta)}
\Big(\frac{\sqrt{(\frac{\rho_H^D}{b_D})^2-(\frac{\nu\beta}{2\pi})^2}b_D}{\sqrt{m^\iota\hat
    h^D_{\iota\kappa}m^\kappa}}\Big)^\frac{d}{2}\\K_{\frac{d}{2}}\Big(2\pi b_D
\sqrt{(\frac{\rho_H^D}{b_D})^2-(\frac{\nu\beta}{2\pi})^2}\sqrt{m^\iota\hat
  h^D_{\iota\kappa}m^\kappa}\Big),
\end{multline}
and
\begin{multline}
  \frac{V_{d-q}}{( \text{V}^D_{q})^{\frac{d-q}{q}}}\sideset{}{'}\sum_{m^\iota\in
    \mathbb Z^q}
\sum_\pm\sum_{n\in \mathbb
  N^*}2 b_D^{\frac d2}\Big(\frac{\sqrt{(\frac{\rho_H^D}{b_D})^2+(n\pm i \frac{\nu\beta}{2\pi})^2}}{\sqrt{m^\iota\hat
    h^D_{\iota\kappa}m^\kappa}}\Big)^\frac{d}{2}\\K_{\frac{d}{2}}\Big(2\pi b_D
\sqrt{(\frac{\rho_H^D}{b_D})^2+(n\pm i \frac{\nu\beta}{2\pi})^2}\sqrt{m^\iota\hat
h^D_{\iota\kappa}m^\kappa}\Big)e^{2\pi im^\iota N_\iota^Dn\mp m^\iota N_\iota^D\nu\beta}.
\end{multline}

\section{Relation between sums of Bessel functions}
\label{sec:relat-betw-bess}

From the equality of the low and high temperature expressions \eqref{eq:37a} and
\eqref{eq:37c}, an identity between sums of Bessels functions holds.

In particular, for one small spatial dimension $q=1$, vanishing chemical
potential for linear momentum $N^\iota=0$, we have
\begin{equation}
{b_D=b^{-1}={L_d\over \beta}, \quad \rho_H^D={\rho_H}={\beta L_d\over 2\pi},\quad \hat h_{dd}^{D}=1,\quad  \text{V}_1^D=\text V_1={ L_d}}.
\end{equation}
Using \eqref{BesselK limit}, this identity reduces in the massless limit
$ \rho_H\rightarrow 0^+ $ to
\begin{multline}
\label{eq:bessel iden high low massless}
b
	{\pi }^{-\frac{d+1}{2}}\zeta\big( {d+1\over 2} \big)\Gamma\big( {d+1\over 2} \big)
	{+b^{1-d}
		\sum_{l\in \mathbb{N}^*}(\pi l^2)^{-{d\over 2}}
		\cosh{(l\beta\nu)}}\Gamma\big ( {d\over 2}\big )\\
+4b^{1-{d\over 2}}
	\sum_{l\in \mathbb{N}^*}\sum_{n\in
		\mathbb{N}^*}\big(\frac{n}{l}\big)^{\frac{d}{2}}
	K_{\frac{d}{2}}\big(2\pi l b n\big)\cosh{(l\beta\nu)}
\\
=
b^{-d}\sum_{l\in \mathbb{N}^*}\big({\pi l^2}\big)^{-\frac{d+1}{2}}\cosh(l\nu\beta)
	+4\sum_{l\in
		\mathbb{N}^*}
	\big(\frac{{i\nu L_d}}{2\pi l}\big)^\frac{d}{2}K_{\frac{d}{2}}\big(
	i\nu L_d l \big)\\
+2b^{-\frac{d}{2}}\sum_\pm\sum_{l\in
		\mathbb{N}^*}\sum_{n\in \mathbb
		N^*}
	\big(\frac{{(n\pm i \frac{\nu\beta}{2\pi})}}{l}\big)^\frac{d}{2}
	K_{\frac{d}{2}}\big(2\pi b^{-1}l
	{(n\pm i \frac{\nu\beta}{2\pi}})\big).
\end{multline}
This result was used in an essential way in \cite{Aggarwal:2022rrp}.

\section{Expressions in terms of polylogarithms}
\label{sec:expr-terms-polyl}

In the literature, the results on the partition function of a charged complex
scalar field with no small spatial dimensions, $q=0$, are traditionally
expressed in terms of polylogarithms (see e.g.~\cite{Haber:1981tr} and the
appendix of \cite{Haber:1981ts}).

In this case, it follows from \eqref{eq:69}, the absence of a Casimir energy,
$E_0^{d,0}=0$ and \eqref{eq:7a} that
\begin{equation}
  \label{eq:32}
  \ln Z_{d,0}(\beta,m,\nu)=-
  \frac{V_{d}}{2^{d-1}\pi^{\frac{d}{2}}\Gamma(\frac{d}{2})}
  \sum_\pm I_{d}(\beta,m,\pm\nu),
\end{equation}
with
\begin{equation}
  \label{eq:33}
  I_{d}(\beta,m,\pm\nu)=\int_0^{\infty}\text{d}k\,
  k^{d-1}\ln[1-e^{-\beta(\omega_{k}\mp\nu)}],\quad \omega_{k}=\sqrt{k^2+m^2}.
\end{equation}
After the change of variables, $x=\beta k$, $\bar m=\beta m=2\pi\rho_L$,
$r=\frac{\nu}{m}$, (so that $r\bar m=\nu\beta$, with $\bar m\geq 0$ and
$|r|\leq 1$), the integral becomes
\begin{multline}
  \label{eq:34}
  I_{d}(\beta,m,\pm\nu)=\frac{1}{\beta^d}\int_0^{\infty}{d}x\,
  x^{d-1}\ln[1-e^{-(\sqrt{x^2+\bar m^2}\mp r\bar m)}]\\=-\frac{1}{d\beta^d}
  \int^\infty_0dx \frac{x^{d+1}}{\sqrt{x^2+\bar m^2}(e^{\sqrt{x^2+\bar m^2}\mp r\bar m}-1)}.
\end{multline}
In terms of the integrals defined in \cite{Haber:1981tr,Haber:1981ts},
\begin{equation}
  \label{eq:36}
  \begin{split}
    h_d(\bar m,r)&=\frac{1}{\Gamma(d)}
                   \int^\infty_0dx \frac{x^{d-1}}{\sqrt{x^2+\bar m^2}
                   (e^{\sqrt{x^2+\bar m^2}- r\bar m}-1)},\quad H_d(\bar m,r)=\sum_{\pm}h_d(\bar m,\pm r)\\
    g_d(\bar m,r)&=\frac{1}{\Gamma(d)}
                   \int^\infty_0dx x^{d-1}\frac{1}{e^{\sqrt{x^2+\bar m^2}- r\bar m}-1},\quad G_d(\bar m,r)=\sum_{\pm}\pm g_d(\bar m,\pm r),
  \end{split}
\end{equation}
the partition function may be written as
\begin{equation}
  \label{eq:37}
  \ln Z_{d,0}(\beta,m,\nu)=\frac{V_d}{\beta^d}
  \frac{2\Gamma(\frac{d+3}{2})}{\pi^{\frac{d+1}{2}}}
  H_{d+2}(\bar m,r),
\end{equation}
where the reduplication formula for the Gamma function has been used. The charge
density becomes
\begin{equation}
  \label{eq:38}
  \delta_{d,0}=\frac{\Gamma(\frac{d+1}{2})}{\pi^{\frac{d+1}{2}}\beta^d}G_d(\bar m,r).
\end{equation}

By the change of variables $w=e^{\bar m-\sqrt{x^2+\bar m^2}}$ and a suitable
expansion followed by integration, it is shown in
\cite{Haber:1981tr,Haber:1981ts} that, if
\begin{equation}
  \label{eq:84}
  y=(1- r)\bar m,
\end{equation}
\begin{multline}
  \label{eq:70}
  g_d=\frac{\Gamma(\frac d2)}{\Gamma(d)}\sum^\infty_{k=0}\frac{1}{\Gamma(\frac d2 -k)k!}
  \big[\frac{1}{2\bar m}\big]^{k+1-\frac d2}\big[\bar m\Gamma(\frac d2 +k){\rm Li}_{k+\frac d2}
  (e^{-y})\\+\Gamma(\frac d2 +k+1){\rm Li}_{k+\frac d2+1}
  (e^{-y})\big],
\end{multline}
and
\begin{equation}
  \label{eq:82}
  h_d=\frac{\Gamma(\frac d2)}{\Gamma(d)}\sum^\infty_{k=0}\frac{\Gamma(\frac d2+k)}{\Gamma(\frac d2 -k)k!}
  \big[\frac{1}{2\bar m}\big]^{k+1-\frac d2}{\rm Li}_{\frac d2+k}
  (e^{-y}),
\end{equation}
where the sums cut for even $d$ at $k=\frac d2 -1$. These expressions are
adapted to a low temperature expansion, $\bar m\gg 1$. To leading order,
\begin{equation}
  \label{eq:83}
  g_d\approx\frac{\Gamma(\frac d2)2^{\frac d2 -1}}{\Gamma(d)}\bar m^{\frac d2}{\rm Li}_{\frac d2}(e^{-y}),\quad
  h_d\approx \frac{\Gamma(\frac d2)2^{\frac d2 -1}}{\Gamma(d)}\bar m^{\frac d2-1}{\rm Li}_{\frac d2}(e^{-y}).
\end{equation}
There are thus two expansions adapted to a low temperature expansion, one in
terms of series of Bessel functions and the other in terms of polylogarithms.
For even $d$, the polylog expression is also valid for a high temperature
expansion since the series is finite.

When comparing \eqref{eq:19a} in the absence of small spatial dimensions with
\eqref{eq:37} and \eqref{eq:82}, it follows that
\begin{equation}
  \label{eq:85}
  h_d={1 \over\Gamma({d+1\over 2})}
  \sum_{p=1}^{\infty}e^{r\bar m p}\left(\frac{\bar m}{2p} \right)^{d-1\over 2}K_{d-1\over 2}(\bar m p).
\end{equation}
This is also shown in \cite{Haber:1981tr}, equation (9), directly from the
definition of the integrals. From \eqref{eq:82} and \eqref{eq:85}, the following
relation between (asymptotic) series of Bessel functions and polylogarithms
holds,
  \begin{multline}
    \label{eq:86}
  \sqrt{\frac 2\pi} \sum_{p=1}^\infty e^{(\bar{m}-y)p}
\left (\frac{\bar{m}} p\right )^{\frac {d-1} 2 }K_{\frac {d-1}2 }(\bar{m} p)=\\
\begin{cases}\sum_{k=0}^{\frac d2 -1}(\bar{m})^{\frac d2 -1-k}
\frac{(\frac d2 -1+k)!}{(\frac d2 -1-k)!k!2^k}{\rm Li}_{\frac d2 +k}(e^{-y}), \quad \text{even}\; d,\\
\sum_{k=0}^{\infty}(\bar{m})^{\frac d2 -1-k}
\frac{\Gamma(\frac d2 +k)}{\Gamma(\frac d2 -k)k!2^k}{\rm Li}_{\frac d2 +k}(e^{-y}), \quad \text{odd}\; d.
\end{cases}
\end{multline}
This may be directly proved as follows. For even $ d $, one uses~\eqref{eq:88}
with $n=\frac{d}{2}-1$
to show that the LHS of \eqref{eq:86} is given by
\begin{equation}
	\label{eq:72}
	\sum_{p=1}^\infty e^{-yp}
	p^{1-d}\sum_{l=0}^{\frac d2 -1}\frac{(\frac d2 -1+l)!}{(\frac d2 -1-l)!l!2^l}(\bar{m} p)^{\frac d2 -1-l}.
\end{equation}
Summing over $p$ using \eqref{eq:31} gives
\begin{equation}
	\label{eq:73a}
	\sum_{l=0}^{\frac d2 -1}(\bar{m})^{\frac d2 -1-l}
	\frac{(\frac d2 -1+l)!}{(\frac d2 -1-l)!l!2^l}{\rm Li}_{\frac d2 +l}(e^{-y}).
\end{equation}
For odd $d$, one uses instead \eqref{eq:87} at $\nu=\frac{d-1}{2}$.

For a high temperature expansion,  $\bar m \ll 1$, one cannot expand as a power series
around $\bar m=0$ because the integrand of $h_d$ has a branch point.
When using that
\begin{equation}
  \label{eq:19e}
  \begin{split}
    &\frac{\partial}{\partial \bar m}\big(\frac{1}{e^{\sqrt{x^2+\bar m^2}\mp r\bar
        m}-1}\big) =(\frac{\bar m}{x}\mp\frac{r\sqrt{x^2+\bar m^2}}{x})
    \frac{\partial}{\partial x}\big(\frac{1}{e^{\sqrt{x^2+\bar m^2}\mp r\bar m}-1}\big),\\
    &\frac{\partial}{\partial \bar r}\big(\frac{1}{e^{\sqrt{x^2+\bar m^2}\mp r\bar
        m}-1}\big)=\mp\frac{\bar m}{x}\sqrt{x^2+\bar m^2}\frac{\partial}{\partial
      x}\big(\frac{1}{e^{\sqrt{x^2+\bar m^2}\mp r\bar m}-1}\big),
  \end{split}
\end{equation}
one shows by integrations by parts the recursion relations
\begin{equation}
  \label{eq:20e}
  \begin{split}
    &\frac{\partial}{\partial \bar m} g_{d+1}=drh_{d+1}(\bar m,r)-\frac{\bar m}{d}g_{d-1}
    +\frac{\bar m^2r}{d}h_{d-1},\\
    &\frac{\partial}{\partial \bar m} h_{d+1}=\frac{r}{d}g_{d-1}-\frac{\bar m}{d}h_{d-1},\\
    &\frac{\partial}{\partial \bar r} g_{d+1}=\bar m d h_{d+1}+\frac{\bar m^3}{d}h_{d-1},\\
    &\frac{\partial}{\partial \bar r} h_{d+1}=\frac{\bar m}{d}g_{d-1},
  \end{split}
\end{equation}
with the same relations holding for $G_{d+1},H_{d+1}$. From the definitions of the
integrals in \eqref{eq:36} and \eqref{eq:89},
the initial conditions are $g_d(0,0)=\zeta(d)$ for $d> 1$,
$h_{d}(0,0)=\frac{\zeta(d-1)}{d-1}$ for $d> 2$
$G_d(0,0)=0$ and $H_d(0,0)=\frac{2\zeta(d-1)}{d-1}$ for
$d>2$.

By the change of variables $t=e^{-\sqrt{x^2+\bar m^2}}$, one finds directly that
\begin{multline}
  \label{eq:21}
  g_2=-\int^{e^{-\bar m}}_0 dt\frac{\ln t}{e^{-r\bar m}-1}={\rm Li}_2[e^{(r-1)\bar m}]-\bar m
  \ln (1-e^{(r-1)\bar m})\\={\rm Li}_2[e^{-y}]+\bar m
  {\rm Li}_1(e^{-y}),
\end{multline}
\begin{equation}
  \label{eq:21a}
  h_2=\int^{e^{-\bar m}}_0 dt\frac{1}{e^{-r\bar m}-1}= -\ln (1-e^{(r-1)\bar m})={\rm Li}_1(e^{-y}),
\end{equation}
Their high temperature expansion is then obtained by using \eqref{eq:38 b}.
Furthermore, these expressions give access to $g_d,h_d,G_d,H_d$ for positive
even $d$ through the recursion relations.

As discussed in Appendix B of \cite{Haber:1981ts}, one may first work out a high
temperature expansion for $G_{1}$ and $H_{1}$ and then use the recursion
formulas to get the high temperature results for odd $d$.

\section{Justification of \eqref{eq:73}}
\label{check}

We will only argue why the expansion of the first line of \eqref{eq:73} is
valid, and assume that the reasoning for the second line is similar. When taking
into account \eqref{eq:85}, the first line of the high-temperature expansion in
\eqref{eq:37c} may be written as
\begin{equation}
  \frac{\text{V}_{d-q}b_D^d}{(\text{ V}^D_q)^{\frac{d-q}q}}
  \frac{2\Gamma(\frac {d+3}{2})}{\pi^{\frac{d+1}{2}}}H_{d+2}(\bar m,r).
\end{equation}
As argued in \cite{Haber:1981tr}, when $\bar m\ll 1$ as in the high
temperature regime \eqref{eq:40a}, then \eqref{eq:85} is not suitable. One needs
to use a series appropriate to this regime which is given by (D1) and (D3) of
\cite{Haber:1981tr}. Using these results one obtains, in the current notation,
\begin{align}
	& H_{d+2}(\bar m,r)= \frac{2\zeta(d+1)} {d+1}
		-\frac{\bar m^2}{(d-1)(d+1)}\zeta(d-1)  (1+(1-d)r^2)+\mathcal O (\bar m^3), \ d>2,\\
		& H_{4}(\bar m,r)= \frac{2\zeta(3)} {3} +\frac{\bar m^2(1-r^2)}{6}\left[\ln (\bar m^2(1-r^2))-\gamma-\psi(2)-2r^2 \right]+\mathcal O (\bar m^3),\
		 \\
		 	& H_{3}(\bar m,r)= {\zeta(2)} -\frac{\pi \bar m}{2}\sqrt{1-r^2}+\mathcal O (\bar m^2).
\end{align}
Here, we used
\begin{equation}
_2F_1(0,b;c;r^2)=1,\quad _2F_1(-1,b;c;r^2)=1-\frac{b}{c} r^2, \quad {}_3F_2(a_1,a_2,0;b_1,b_2;r^2)=1~.
\end{equation}
To compute the charge density, we need
\begin{align}
	\partial_\nu H_{d+2}(\bar m,r)=\frac{1}{m}\partial_r H_{d+2}(\bar m,r),
\end{align}
and thus
\begin{align}
\partial_\nu	H_{d+2}(\bar m,r)&=
	\frac{2\beta^2\nu}{(d+1)}\zeta(d-1)  +\mathcal O (\bar m^2)~, \quad d>2\\
\partial_\nu	H_{4}(\bar m,r)&=  -\frac{\beta^2\nu}{3}\left[\ln (\bar m^2(1-r^2))-4r^2+3-\gamma-\psi(2) \right]
+\mathcal O (\bar m^2),
	\\
\partial_\nu	H_{3}(\bar m,r)&= \frac{\pi \beta r	}{2\sqrt{1-r^2}}+\mathcal O (\bar m^1).
\end{align}
It follows that, for $d>2$, the first line of \eqref{eq:68} becomes
\begin{equation}
 \frac{b_D^d\beta\nu}{( \text{V}_{q}^D)^{\frac{d-q}{q}}}
	\Big[
	\frac{	2}{\pi^{\frac{d+1}{2}}}\zeta(d-1)\Gamma (\dfrac{d+1	}{2} )++\mathcal O (\bar m^2)\Big],
\end{equation}
which matches the first term of \eqref{eq:73} and thus
\eqref{eq:critical density}. For $d=1,2$, the critical density at $r=1$
diverges, so there is no Bose-Einstein condensation.

\addcontentsline{toc}{section}{References}

\printbibliography

\end{document}